\documentclass[fleqn,usenatbib,useAMS]{mnras}

\usepackage[T1]{fontenc}
\usepackage{ae,aecompl}

\usepackage{newtxtext,newtxmath}

\usepackage{graphicx}	    
\usepackage{amsmath}	    
\usepackage{amssymb}	    
\usepackage{multicol}       
\usepackage{bm}		        
\usepackage{pdflscape}	    
\usepackage{multirow}       
\usepackage{lineno}         
\usepackage{xcolor}         
\usepackage{booktabs}

\graphicspath{{figures/}}

\newcommand{\gjoll}{Gj\"{o}ll}

\input{hyperlink-year-only-natbib-patch}

\title[The VPOS revisited]{The Milky Way's stellar streams and globular clusters do not align in a Vast Polar Structure}

\author[A.~H.~Riley \& L.~E.~Strigari]{
Alexander H.~Riley$^{1,2}$\thanks{E-mail: \href{mailto:alexriley@tamu.edu}{alexriley@tamu.edu}. Code for this work is available \href{https://github.com/ahriley/vpos-revisit}{on Github}.}\href{https://orcid.org/0000-0001-5805-5766}{\includegraphics[scale=0.1]{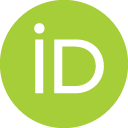}} and Louis E.~Strigari$^{1,2}$
\\
$^{1}$George P. and Cynthia Woods Mitchell Institute for Fundamental Physics and Astronomy, Texas A\&M University, College Station, TX 77843, USA\\
$^{2}$Department of Physics and Astronomy, Texas A\&M University, College Station, TX 77843, USA}

\date{Accepted 2020 March 9. Received 2020 March 6; in original form 2020 January 30}

\pubyear{2020}

\begin{document}
\label{firstpage}
\pagerange{\pageref{firstpage}--\pageref{lastpage}}
\maketitle

\begin{abstract}
There is increasing evidence that a substantial fraction of Milky Way satellite galaxies align in a rotationally-supported plane of satellites, a rare configuration in cosmological simulations of galaxy formation.
It has been suggested that other Milky Way substructures (namely young halo globular clusters and stellar/gaseous streams) similarly tend to align with this plane, accordingly dubbed the Vast Polar Structure (VPOS).
Using systemic proper motions inferred from {\it Gaia} data, we find that globular cluster orbital poles are not clustered in the VPOS direction, though the population with the highest VPOS membership fraction is the young halo clusters ($\sim$30\%).
We additionally provide a current census of stellar streams, including new streams discovered using the Dark Energy Survey and {\it Gaia} datasets, and find that stellar stream normals are also not clustered in the direction of the VPOS normal.
We also find that, based on orbit modeling, there is a likely association between NGC 3201 and the Gj\"{o}ll stellar stream and that, based on its orbital pole, NGC 4147 is likely not a Sagittarius globular cluster.
That the Milky Way's accreted globular clusters and streams do not align in the same planar configuration as its satellites suggests that the plane of satellites is either a particularly stable orbital configuration or a population of recently accreted satellites.
Neither of these explanations is particularly likely in light of other recent studies, leaving the plane of satellites problem as one of the more consequential open problems in galaxy formation and cosmology.
\end{abstract}

\begin{keywords}
galaxies: kinematics and dynamics -- Local Group -- galaxies: formation
\end{keywords}


\section{Introduction} \label{sec:intro}

\begin{figure*}
\includegraphics[width=\linewidth]{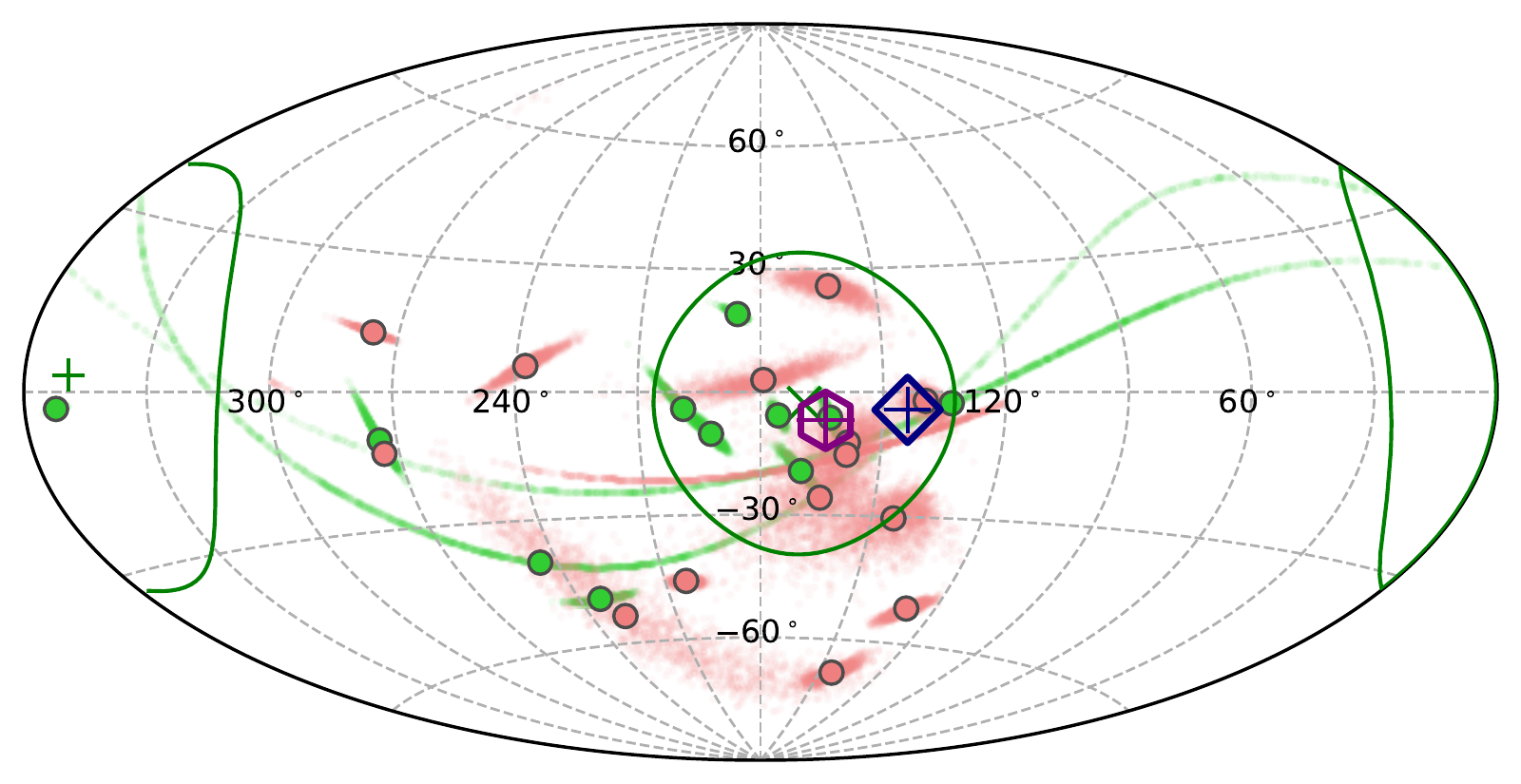}
\caption{All-sky distribution of orbital poles for Milky Way classical satellites (green) and normals for stellar and gaseous streams discovered prior to 2012 \citep[][red]{Pawlowski:2012} in an Aitoff projection.
We use spherical coordinates $(l_\text{MW},b_\text{MW})$ that are aligned with the heliocentric Galactic system and centered on the Milky Way center \citep{Pawlowski:2020}.
Orbital poles for the satellites are calculated as described in Section \ref{subsec:galactocentric} using data compiled by \citet{Riley:2019}.
Since the orbital sense is unknown for most streams, only one of the two stream normals (those from $120^\circ < l_\text{MW} < 300^\circ$) is shown.
The large points correspond to the observed (assuming no measurement errors) orbital poles, while the point clouds show the results of 2,000 Monte Carlo simulations incorporating measurement errors.
We also show the normal vector of a disk fitted to the spatial distribution of young halo globular clusters \citep[][blue diamond]{Pawlowski:2012} and the minor axis direction of a spatial plane fitted to all Milky Way satellites \citep[][purple hexagon]{Pawlowski:2015}.
The green circles each contain 10\% of the sky around the assumed VPOS pole \citep{Pawlowski:2013, Fritz:2018}, given as an ``x'' for the co-orbiting direction and a ``+'' for the counter-orbiting direction.
{\it The previously observed preferential alignment of Milky Way satellite galaxies, stellar streams, and young halo globular clusters has been dubbed the Vast Polar Structure (VPOS).}}
\label{fig:vpos}
\end{figure*}

For over 40 years, it has been known that the classical\footnote{Throughout this work, ``classical'' refers to the 11 Milky Way satellites discovered prior to the Sloan Digital Sky Survey: Carina, Draco, Fornax, Leo I and II, the Large and Small Magellanic Clouds, Sagittarius, Sculptor, Sextans, and Ursa Minor.} satellites of the Milky Way fall on the same polar great circle as the Magellanic Stream \citep{Lynden-Bell:1976, Kunkel:1976}.
More recent studies have confirmed the existence of the Milky Way's plane of satellites \citep{Lynden-Bell:1982, Kroupa:2005, Metz:2007} and found that the orbital poles of its constituents cluster in the same direction as the spatial plane \citep{Metz:2008, Pawlowski:2013}, establishing the plane of satellites as a stable, co-rotating system.
Similarly thin, kinematically-coherent satellite planes have been observed around M31 \citep{Conn:2013, Ibata:2013, Santos-Santos:2019} and Centaurus A \citep{Muller:2018}, with contested statistical evidence of such systems around more distant host galaxies \citep{Ibata:2014external, Ibata:2015, Cautun:2015, Phillips:2015}.

Particular attention has been paid to the Milky Way's plane of satellites, as it is the system with the most information and highest quality data by nature of our place as observers.
As more (and fainter) satellite galaxies have been discovered in the Sloan Digital Sky Survey, the Dark Energy Survey, and smaller community efforts \citep{Simon:2019}, these new satellites continue to preferentially reside in the same spatial plane \citep{Metz:2009SDSS, PawlowskiKroupa:2014, Pawlowski:2015}.
Proper motion measurements using data from {\it Gaia} Data Release 2 \citep[DR2;][]{GaiaDR2:2018} show that the orbital poles for non-classical satellites also preferentially cluster in this plane \citep{Fritz:2018}. 
Overall, 19 of the 41 satellites with 6-D phase-space measurements have orbital poles that align with the VPOS, while an additional 10 have insufficient proper motion accuracy to establish whether they align.
Furthermore, the degree of kinematic coherence for the classical satellites has increased as measurements continue to improve, as expected if the underlying distribution was truly correlated \citep{Pawlowski:2020}.

Most notably, \citet{Pawlowski:2012} found that Milky Way stellar/gaseous streams and the spatial distribution of young halo globular clusters preferentially align with the satellite galaxy plane.
This dramatically strengthened the case for highly-correlated structure formation in the Milky Way, connecting three different sets of accreted objects.
Since this finding, the Milky Way's planar structure has usually been referred to as the Vast Polar Structure (VPOS), comprised of satellite galaxies, globular clusters, and stellar streams.

Satellite planes as thin and kinematically correlated as observed in the Milky Way are extremely rare in cosmological simulations \citep{Ibata:2014, Pawlowski:2014, Cautun:2015simscompare, Shao:2019}, posing a small-scale challenge to our current $\Lambda$CDM cosmological paradigm \citep{BullockBK:2017}.
All proposed solutions within $\Lambda$CDM have, thus far, not been able to explain such kinematic coherence satisfactorily, including: infall of satellites in groups \citep{DOnghia:2008, Metz:2009, Wang:2013, Shao:2018}, accretion along a cosmic filament \citep{Zentner:2005, Lovell:2011, Libeskind:2011, Pawlowski:2014}, the inclusion baryonic effects in cosmological simulations \citep{Pawlowski:2015baryons, Ahmed:2017, Muller:2018, Pawlowski:2019PHAT}, and special environments or properties of the host halos \citep{Pawlowski:2014ELVIS, Buck:2015, Pawlowski:2019PHAT}.
For further discussion on the observations of satellite planes, comparisons to cosmological simulations, and possible solutions to the problem, see \citet{Pawlowski:2018}.

In light of the challenge that the VPOS poses to our theory of hierarchical structure formation, it is worthwhile to re-visit the analysis in \citet{Pawlowski:2012} as our knowledge and understanding of the Milky Way's substructures has evolved.
In particular, there have been many newly discovered streams using high-quality datasets, like those discovered in the Dark Energy Survey \citep{Shipp:2018} and in {\it Gaia} DR2 with the \texttt{STREAMFINDER} algorithm \citep{Malhan:2018, Ibata:2019}.
While some new streams have been analyzed in the context of the VPOS \citep{PawlowskiKroupa:2014, Grillmair:2017a, Grillmair:2017b, Shipp:2018}, a consistent analysis of all recently discovered streams has been lacking to this point.

In addition, courtesy of {\it Gaia} DR2, we now have 6-D phase space measurements for nearly every Milky Way globular cluster \citep{GaiaHelmi:2018, Vasiliev:2019}.
This enables an assessment of VPOS membership for individual globular clusters, similar to the calculation performed in \citet{Fritz:2018} for satellite galaxies, rather than analyzing the spatial distribution of the globular cluster system as a whole \citep{Pawlowski:2012, Arakelyan:2018}.
If a significant fraction of individual globular clusters have orbital poles that align with the VPOS, it would strengthen the case for the VPOS even further.

It is not immediately clear if all of this new data confirms previous results from \citet{Pawlowski:2012}, strengthening the tension with $\Lambda$CDM, or possibly confuses the interpretation of the planes of satellites problem.
To address this, we revisit the alignment of Milky Way stellar streams and globular clusters with the VPOS in light of recent discoveries and measurements.
In particular, we calculate globular cluster orbital poles (Section \ref{sec:globulars}) and stream normals (Section \ref{sec:streams}), evaluating the likelihood that they are members of the VPOS while taking observational errors into account.
We also employ orbit modeling (Section \ref{sec:orbits}) in an attempt to associate Milky Way satellites with known streams.
We discuss these results in the context of other recent studies in Section \ref{sec:discussion} and summarize our findings in Section \ref{sec:summary}.

\section{Globular Clusters} \label{sec:globulars}

In this section, we analyze the globular cluster system of the Milky Way for preferential alignment of orbital poles with the VPOS normal.
We divide the system into three populations according to their expected origin: old halo clusters that formed with the early Milky Way, young halo clusters that were accreted over time, and bulge/disk clusters that are confined to the Milky Way's bulge and disk.

The accretion origin of the young halo clusters makes them particularly interesting in the context of the plane of satellites, as satellite galaxies also formed outside of the Milky Way and were accreted over time.
If the young halo clusters have a similar spatial and kinematic distribution to the planar satellite galaxies, this would strengthen the case that the plane of satellites is both real and tied to the formation history of the Milky Way.
Indeed, the young halo clusters have been found to have a flattened spatial distribution that aligns with the VPOS normal \citep{Pawlowski:2012, Arakelyan:2018}.
Despite this finding, there has not been a detailed analysis of globular cluster orbital poles in the context of the VPOS.\footnote{
\citet{Vasiliev:2019} briefly mentioned that the full system of globular cluster orbital poles does not appear to cluster with the VPOS normal. 
However, analyzing the globular clusters as one system mitigates any expected VPOS signal.
For example, bulge/disk clusters are expected have orbital poles that tend to align with the Galactic poles, which are offset $\sim$90$^\circ$ from the VPOS.
}
We aim to rectify this situation here.

We use the classifications presented in \citet{Mackey:2005} that are based on metallicity and horizontal-branch morphology index.
This results in 70 old halo clusters, 30 young halo clusters, and 35 bulge/disc clusters, as well as 9 clusters that have been discovered more recently and were not classified in that work.
We also analyze 6 clusters that were thought to be associated with the Sagittarius dwarf galaxy, though these are not our focus as Sagittarius's orbital pole is almost perfectly misaligned with the VPOS normal.

\subsection{Computing Galactocentric quantities} \label{subsec:galactocentric}

The 6-D phase-space information for each globular cluster is taken from the \citet{Vasiliev:2019} catalog of proper motions derived using data from {\it Gaia} Data Release 2 \citep{GaiaDR2:2018}.
The catalog uses spatial information from \citet{Harris:1996} (2010 edition) and radial velocity measurements from \citet{Baumgardt:2019}.
As in \citet{Vasiliev:2019}, we assume an error of 0.1 in distance modulus for each cluster, corresponding to a relative error of 0.046 in the distance.

The orbital poles are computed relative to the center of the Milky Way as described in \citet{Pawlowski:2013}.
In summary, we convert each object's heliocentric position, radial velocity, and proper motion into a Galactocentric Cartesian frame and calculate the angular momentum vector as the cross product of the position and velocity vectors. We then decompose this vector into a normalized direction (orbital pole).
The orbital poles are reported in a spherical coordinate system ($l_\text{MW}$, $b_\text{MW}$) that is centered on the Milky Way center and aligned with the standard heliocentric Galactic frame.

To estimate uncertainties, we perform 2,000 Monte Carlo simulations drawing from the uncertainties in the distance, radial velocity, and proper motions of each cluster.
When drawing the proper motions, we account for the correlation between proper motions $\mu_{\alpha*} \equiv \mu_\alpha\cos\delta$ and $\mu_\delta$.

We assume a Galactocentric frame with a distance from the Sun to the Galactic centre of 8.122~kpc \citep{GRAVITY:2018}, a height of the Sun relative to the local Galactic mid-plane of $20.8$~pc \citep{Bennett:2019}, and a solar motion relative to the Galactic centre of $(v_x, v_y, v_z)$ = (12.9, 245.6, 7.78)~km~s$^{-1}$ (\citealt{Drimmel:2018}, who combine exquisite measurements from \citealt{GRAVITY:2018} and \citealt{Reid:2004}), where the $x$-axis points from the position of the Sun projected on to the Galactic plane to the Galactic centre, the $y$-axis points towards Galactic longitude $l=90^\circ$ (i.e.~in the direction of Galactic rotation), and the $z$-axis points towards the North Galactic Pole.
We note that, as for satellite galaxies \citep{Metz:2008, Pawlowski:2013}, the orbital pole uncertainties for the globular clusters are dominated by uncertainties in the proper motion measurements.
We therefore we refrain from incorporating the (small) uncertainties on Galactocentric frame properties in our analysis.

\subsection{VPOS membership} \label{subsec:GCVPOS}

\begin{figure*}
\includegraphics[width=\linewidth]{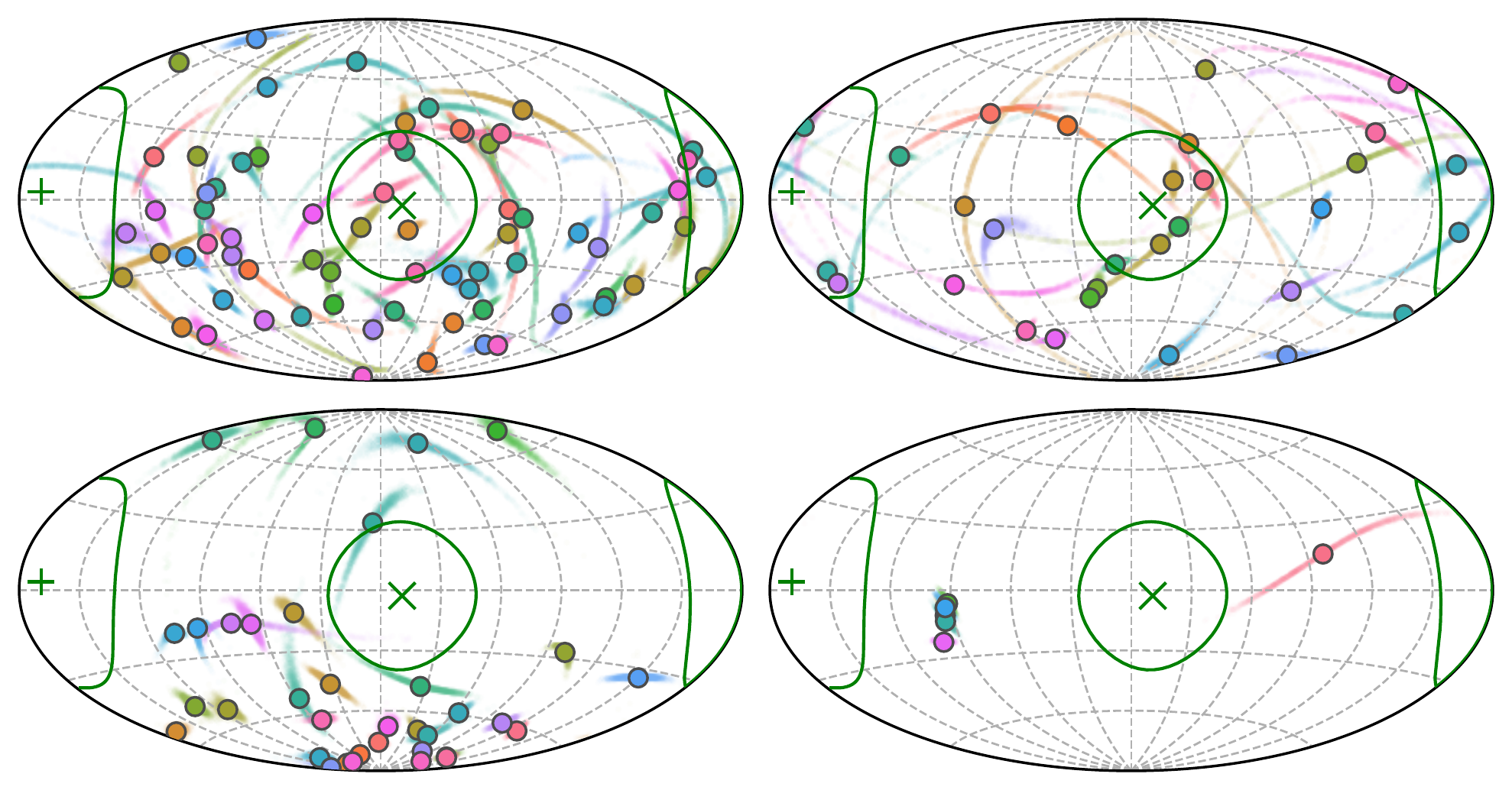}
\caption{All-sky distribution of orbital poles for Milky Way globular clusters.
The large points correspond to the observed (assuming no measurement errors) orbital poles, while the point clouds show the results of 2,000 Monte Carlo simulations incorporating measurement errors.
The green circles each contain 10\% of the sky around the assumed VPOS pole, given as an ``x'' for the co-orbiting direction and a ``+'' for the counter-orbiting direction.
Each panel corresponds to a different population of globular clusters using the classifications from \citet{Mackey:2005}: old halo (top left), young halo (top right), bulge/disc (bottom left), and Sagittarius (bottom right).
{\it Only the young halo cluster population includes a substantial fraction of VPOS members ($\sim$30\%, see Table \ref{tab:GCsummary}).}}
\label{fig:GCpoles}
\end{figure*}

Figure \ref{fig:GCpoles} shows the resulting distribution of globular cluster orbital poles, broken down by the different populations.
The observed orbital pole (assuming no error in measurements) of each cluster is shown as a large circle and the uncertainty of the orbital pole is illustrated by the point clouds based on 2,000 Monte Carlo realizations.
The green cross indicates the assumed VPOS normal of $(l_\text{MW}, b_\text{MW}) = (169.3^\circ, -2.8^\circ)$, as used to predict satellite galaxy proper motions in \citet{Pawlowski:2013} and evaluate their VPOS membership in \citet{Fritz:2018}.
The green circles of opening angle $\theta_\text{inVPOS} = 36.87^\circ$ denote areas of 10\% of the sphere around the VPOS normal.
Orbital poles that lie within this region are considered to orbit along the VPOS \citep{Fritz:2018}.

Based on the assumption that the cluster orbits along the VPOS, we can find the predicted orbital pole direction that best aligns with the VPOS based on spatial information alone.
This is the direction along a great circle oriented perpendicular to the position of the cluster (as seen from the center of the Milky Way) which minimizes the angle to the assumed VPOS normal.
With this optimal orbital pole, we can calculate the same summary statistics as in \citet{Fritz:2018} to evaluate if a cluster is a member of the VPOS:
\begin{itemize}
    \item $\theta_\text{pred}$: the angle between the predicted orbital pole and the VPOS normal.
    If this exceeds $\theta_\text{inVPOS}$, then VPOS membership is ruled out based on spatial information alone
    \item $\theta_\text{obs}$: the angle between the observed orbital pole and the VPOS normal
    \item $p_\text{inVPOS}$: the fraction of Monte Carlo sampled orbital pole directions that fall within $\theta_\text{inVPOS}$ of the VPOS normal.
    This estimates the probability that a cluster is a VPOS member
    \item $p_\text{>VPOS}$: the fraction of Monte Carlo orbital pole directions that are further from the VPOS normal than $\theta_\text{inVPOS}$ {\it after artificially rotating the Monte Carlo samples} so that the observed pole falls on the predicted pole.
    This estimates the probability of falsely finding the cluster's orbital pole to be misaligned with the VPOS
    \item $p_\text{>obs}$: the fraction of Monte Carlo orbital pole directions that are further from the VPOS normal than $\theta_\text{obs}$ {\it after artificially rotating the Monte Carlo samples} so that the observed pole falls on the predicted pole.
    This estimates the probability of measuring an intrinsically well-aligned orbital pole as far away from the VPOS normal as observed. As this procedure assumes intrinsically aligned poles, this value should be considered a lower limit
\end{itemize}

A good candidate for VPOS membership would have $\theta_\text{obs} < \theta_\text{inVPOS}$, a large $p_\text{inVPOS}$, and a small $p_\text{>VPOS}$.
We consider systems with $p_\text{inVPOS} > 0.5$ to be VPOS members, $p_\text{inVPOS} < 0.05$ to firmly and conclusively rule out VPOS membership, and intermediate values as too weakly constrained to be conclusive.
This scheme is simple and reproduces most of the classifications that \citet{Fritz:2018} arrived at for satellite galaxies (they do not state their classification criteria explicitly).

\subsection{Results and comparison to isotropy}

\begin{table*}
\centering
\begingroup
\renewcommand{\arraystretch}{1.25}
\begin{tabular}{lcccccrccc}
	\toprule
	Name & Type & $l_\text{pole}$ & $b_\text{pole}$ & $\Delta_\text{pole}$ & $\theta_\text{pred}$ & $\theta_\text{obs}$ & $p_\text{inVPOS}$ & $p_\text{>VPOS}$ & $p_\text{>obs}$ \\
	 &  & [deg] & [deg] & [deg] & [deg] & [deg] &  &  &  \\
	\midrule
	NGC 104 & BD & 52.5 & -61.7 & 1.4 & 51.1 & $-80.2^{+0.1}_{-0.1}$ & 0.000 & 0.000 & 0.000 \\
	NGC 288 & OH & 133.2 & 32.6 & 8.1 & 44.7 & $49.2^{+4.2}_{-3.0}$ & 0.000 & 0.002 & 0.000 \\
	NGC 362 & YH & 143.7 & 9.6 & 15.7 & 27.0 & $29.1^{+2.2}_{-0.9}$ & 0.946 & 0.040 & 0.065 \\
	Whiting 1 & UN & 284.1 & -20.8 & 5.5 & 43.7 & $-67.9^{+3.8}_{-3.4}$ & 0.000 & 0.000 & 0.000 \\
	NGC 1261 & YH & 265.2 & 39.7 & 17.2 & 21.6 & $-79.4^{+7.2}_{-14.0}$ & 0.004 & 0.038 & 0.000 \\
	\bottomrule
\end{tabular}
\endgroup
\caption{Alignment with the VPOS for individual globular clusters (abridged version, see Table \ref{tab:GCprops-full} for full list).
`Type' refers to the origin of the cluster \citep{Mackey:2005}: old halo (OH), young halo (YH), bulge/disc (BD), Sagittarius (SG), and unclassified (UN).
The orbital pole is reported in spherical coordinates $(l_\text{MW},b_\text{MW})$ that are aligned with the heliocentric Galactic system and centered on the Milky Way center.
$\Delta_\text{pole}$ is the angular uncertainty on the orbital pole direction, computed using the spherical standard distance.
$\theta_\text{pred}$ is the minimum possible angle between the VPOS normal and the cluster's orbital pole based on spatial information alone. $\theta_\text{obs}$ is the observed angle (positive value indicates co-orbiting with respect to the VPOS, negative for counter-orbiting).
$p_\text{inVPOS}$ is the probability that the cluster's orbital pole aligns with the VPOS normal to within the 10\% circle ($\theta_\text{inVPOS} = 36.87^\circ$).
$p_\text{>VPOS}$ is the probability of falsely finding an intrinsically perfectly aligned orbital pole outside of this area given the measurement uncertainties.
$p_\text{>obs}$ is the probability of finding an orbital pole at least as far inclined from the VPOS as the observed orbital pole.
See Section \ref{subsec:GCVPOS} for details on these calculations.
A likely VPOS member would have $\theta_\text{obs} < \theta_\text{inVPOS}$, a large $p_\text{inVPOS}$, and a small $p_\text{>VPOS}$.}
\label{tab:GCprops}
\end{table*}

Table \ref{tab:GCprops} provides an overview of the statistics used to determine VPOS membership.
For the total population of Milky Way globular clusters, there is no obvious clumping of orbital poles within the confines of the VPOS; according to the criteria from Section \ref{subsec:GCVPOS}, 19 clusters are VPOS members, 113 have well-constrained orbital poles that are misaligned with the VPOS, and 18 are not constrained enough to tell.
These results are broadly consistent with \citet{Vasiliev:2019}.

\begin{table}
\centering
\begin{tabular}{cccccc}
	\toprule
	Type & $N$ & $f_\text{inVPOS}$ & $f_\text{notVPOS}$ & $f_\text{inconclusive}$ & $\sum p_\text{inVPOS} / N$ \\
	\midrule
	OH & 70 & 0.143 & 0.743 & 0.114 & 0.130 \\
	YH & 30 & 0.300 & 0.467 & 0.233 & 0.327 \\
	BD & 35 & 0.000 & 0.971 & 0.029 & 0.011 \\
	SG & 6 & 0.000 & 1.000 & 0.000 & 0.005 \\
	UN & 9 & 0.000 & 0.778 & 0.222 & 0.080 \\
	\midrule
	All & 150 & 0.127 & 0.753 & 0.120 & 0.134 \\
	\midrule
	F18 & 41 & 0.463 & 0.293 & 0.244 & 0.437 \\
	\bottomrule
\end{tabular}

\caption{Summary of VPOS membership for each population of Milky Way globular clusters: old halo (OH), young halo (YH), bulge/disc (BD), Sagittarius (SG), and unclassified (UN).
$N$ is the total number of clusters for that population, $f_\text{inVPOS}$ is the fraction of clusters with $p_\text{inVPOS} > 0.5$, $f_\text{notVPOS}$ is the fraction of clusters with $p_\text{inVPOS} < 0.05$, $f_\text{inconclusive}$ is the remaining fraction of clusters for which VPOS membership is inconclusive (see Section \ref{subsec:GCVPOS} for details).
The final column is the sum of the membership probabilities $p_\text{inVPOS}$ for each population divided by the total number in that population; this is another way to estimate the fraction of VPOS members.
We also include the results for satellite galaxies as classified by \citet{Fritz:2018}, with $p_\text{inVPOS}$ values taken directly from their Table 4 (we add the LMC and SMC, assuming $p_\text{inVPOS} = 1$ for both).
{\it The only globular cluster population with a substantial fraction of VPOS members is the young halo clusters, but the fraction of satellite galaxies that are VPOS members is considerably higher.}}
\label{tab:GCsummary}
\end{table}

Separating the globular clusters by their suspected origin reveals a more interesting picture, as seen in the summary provided in Table \ref{tab:GCsummary} and illustrated in Figure \ref{fig:GCpoles}. 
The most interesting population with respect to the VPOS is the young halo clusters.
\citet{Pawlowski:2012} found that the young halo clusters have a flattened spatial distribution that is aligned with the Milky Way's plane of satellite galaxies.
We find that the orbital poles of young halo globular clusters are much more likely to align with the VPOS normal than those of other populations.
Specifically, 9 of the 30 clusters have $p_\text{inVPOS} > 0.5$ and the sum of the membership probabilities (another way to estimate the number of VPOS members) for the young halo clusters is $\sum p_\text{inVPOS} = 9.82$.

It can be enlightening to compare this result to expectations from an isotropic distribution.
The co-orbiting and counter-orbiting circles in Figure \ref{fig:GCpoles} each encompass 10\% of the sky, so it would be reasonable to expect $\sim$20\% (in the case of young halo clusters, 6 of 30) of uniformly distributed poles to fall within this region.
We can go further and calculate the probability that at least $k$ out of $n$ uniformly distributed orbital poles fall within the 10\% circles on the sky.
This is essentially a sum over Bernoulli experiments (see \citealt{Pawlowski:2012} for details):
\begin{equation}
    P = \sum_{i=0}^{n-k} \begin{pmatrix} n \\ k + i \end{pmatrix} p^{k+i} (1 - p)^{n-k-i},
\end{equation}
where $p = 0.2$.
The probability that at least $k=9$ out of $n=30$ poles fall within $\theta_\text{inVPOS}$ of the VPOS normal is 12.9\% (for $k=10$ this lowers to 6.1\%).
This is considerably higher than the results for satellite galaxies ($k=19$ of $n=41$ poles would fall within the VPOS tolerance 0.01\% of the time).

In contrast, the orbital poles for old halo clusters do not show significant clumping in the direction of the VPOS normal.
Of the 70 old halo clusters, 10 have $p_\text{inVPOS} > 0.5$ resulting in a $f_\text{inVPOS} = 0.143$, approximately half that of the young halo clusters.
The sum of the membership probabilities for old halo clusters is $\sum p_\text{inVPOS} = 9.12$.
The probability that at least $k=10$ out of $n=70$ uniformly distributed poles fall within the VPOS tolerance is 91.5\%.

The bulge/disc clusters behave as expected, with orbital poles tending to fall near the co-orbiting pole of the Milky Way ($b_\text{MW} = -90^\circ$).
Only one of the 35 bulge/disc clusters has non-negligible $p_\text{inVPOS}$ (NGC 6440, with $p_\text{inVPOS} = 0.377$).
These results are consistent with a population of globular clusters forming within the plane of the Milky Way disc (perpendicular to the VPOS normal).

The globular clusters associated with Sagittarius also behave as expected.
Five of the clusters (NGC 6715, Terzan 7, Arp 2, Terzan 8, and Pal 12) have orbital poles that are well constrained and located near $(l_\text{MW}, b_\text{MW}) = (275.2, -8.0)$, the orbital pole of Sagittarius \citep{Pawlowski:2020}.
Four of the five have $\theta_\text{pred} > \theta_\text{inVPOS}$; they are ruled out as VPOS members based on spatial information alone, as is the case with Sagittarius (they correspondingly have $p_\text{inVPOS} = 0$).
The fifth cluster, Pal 12, has $\theta_\text{pred} = 3.34^\circ$ but also $p_\text{inVPOS} = 0$. 
The sixth cluster, NGC 4147, has an orbital pole that is not constrained well but actually nearly counter-orbits relative to Sagittarius (its orbital pole point cloud is clearly offset from those of other Sagittarius clusters in the lower right panel of Figure \ref{fig:GCpoles}).
We discuss this cluster as it relates to Sagittarius in further detail in Section \ref{sec:discussion}, but it is also ruled out as a VPOS member ($p_\text{inVPOS} = 0.023$).

\section{Stellar Streams} \label{sec:streams}

In this section, we analyze the stellar stream system of the Milky Way for preferential alignment with the VPOS normal.
Stellar streams form as initially self-bound objects that fall into the Milky Way and are unraveled by the tidal forces of the Milky Way's potential.
This makes them extremely useful for tracing the hierarchical formation of the Milky Way \citep{Peebles:1965, Press:1974, Blumenthal:1984}.

An alignment of stellar streams with the Milky Way's plane of satellite galaxies strengthens the case of the plane's existence and expands its definition to include non-satellite objects as a Vast Polar Structure.
\citet{Pawlowski:2012} found that 7 of the 14 streams that had been discovered at the time aligned with the VPOS normal.
This distribution of stream normals significantly deviates from an isotropic distribution (the probability of finding at least 7 of 14 isotropically-distributed normals within 32$^\circ$ of the VPOS normal is 0.24\%).

Since that work, there have been several new streams discovered using either photometric surveys (e.g.~\citealt{Shipp:2018}) or kinematics from {\it Gaia} (e.g.~\citealt{Ibata:2019}).
In light of these discoveries, we re-examine the question of whether the distribution of stream normals remains preferentially aligned with the VPOS.
To do this, we will calculate stream normals (\S\ref{subsec:normalcalc}) for every known Milky Way stream (\S\ref{subsec:streamcatalog}) and analyze their alignment relative to the VPOS (\S\ref{subsec:streamresults}).

\subsection{Stream normal calculation} \label{subsec:normalcalc}

As in \citet{Pawlowski:2012}, we assume that stellar streams approximately trace the orbit of their progenitors \citep{Odenkirchen:2003}.
Specifically, we assume that the orbital plane of the stream can be approximated by the plane defined by two anchor points along the stream (most commonly the endpoints) and the Milky Way center.
We note that just as the angular momentum vector is not an integral of motion and orbital poles effectively precess and nutate in a non-spherical potential \citep{BinneyTremaine:2008}, stream normals exhibit a similar behavior \citep{Erkal:2016}.
However, since VPOS membership is best defined as an aligned orbital pole to the VPOS normal \citep{Fritz:2018}, we apply this same criteria using stream normals.

Once the two stream anchor points are determined (see \S\ref{subsec:streamcatalog}), we can calculate the position of the stream normal.
We begin by transforming the anchor points into Cartesian Galactocentric coordinates, assuming the same transformation as in Section \ref{subsec:galactocentric}.
We then compute the normal of the plane as the normalized cross product of the two spatial anchor vectors.
Proper motion information (and therefore the orbital direction) is not available for most streams, so the normals that we report are chosen to lie between $120^\circ < l_\text{MW} < 300^\circ$ to match \citet{Pawlowski:2012}.

To estimate uncertainties, we perform 2,000 Monte Carlo simulations drawing from the uncertainties in the angular positions and distances to each anchor point.
When varying the angular positions, we displace the anchor point in a random direction on the sky by an angular offset drawn from a Gaussian with a mean of zero and a standard deviation given by the angular uncertainty as compiled in Table \ref{tab:streamprops}.
The major source of uncertainty in the stream normal direction typically comes from distance uncertainties; these can be $\sim$20\% for streams discovered in photometric surveys using a matched-filter method \citep{Shipp:2018}.

\subsection{Assembling a catalog} \label{subsec:streamcatalog}

\begin{table*}
\centering
\begin{tabular}{lcrrccccccccr}
	\toprule
	Name & Class & RA & Dec & Distance & $\Delta \theta$ & Length & Width & $l_\text{normal}$ & $b_\text{normal}$ & $\theta_\text{obs}$ & $p_\text{inVPOS}$ & Ref. \\
	 &  & [deg] & [deg] & [kpc] & [deg] & [deg] & [deg] & [deg] & [deg] & [deg] &  &  \\
	\midrule
	\multirow{2}{*}{20.0-1} & \multirow{2}{*}{3} & 112.92 & 61.56 & $12.6 \pm 1.3$ & $1.8$ & \multirow{2}{*}{158.6} & \multirow{2}{*}{1.8} & \multirow{2}{*}{270.2} & \multirow{2}{*}{38.0} & \multirow{2}{*}{$76.4^{+8.3}_{-6.0}$} & \multirow{2}{*}{0.000} & \multirow{2}{*}{\citet{Mateu:2018}} \\
	 &  & 273.92 & -43.35 & $28.1 \pm 2.8$ & $1.8$ &  &  &  &  &  &  &  \\
	\multirow{2}{*}{Acheron} & \multirow{2}{*}{1} & 230.00 & -2.00 & $3.8 \pm 0.8$ & $0.5$ & \multirow{2}{*}{36.5} & \multirow{2}{*}{0.4} & \multirow{2}{*}{228.2} & \multirow{2}{*}{-53.6} & \multirow{2}{*}{$70.4^{+6.9}_{-5.1}$} & \multirow{2}{*}{0.000} & \multirow{2}{*}{\citet{Grillmair:2009}} \\
	 &  & 259.00 & 21.00 & $3.5 \pm 0.7$ & $0.5$ &  &  &  &  &  &  &  \\
	\multirow{2}{*}{ACS} & \multirow{2}{*}{1} & 126.40 & -0.70 & $8.9 \pm 0.2$ & $2.1$ & \multirow{2}{*}{65.1} & \multirow{2}{*}{2.1} & \multirow{2}{*}{143.5} & \multirow{2}{*}{-68.0} & \multirow{2}{*}{$67.5^{+0.4}_{-0.4}$} & \multirow{2}{*}{0.000} & \multirow{2}{*}{\citet{Grillmair:2006a}} \\
	 &  & 133.90 & 64.20 & $8.9 \pm 0.2$ & $2.1$ &  &  &  &  &  &  &  \\
	\multirow{2}{*}{Aliqa Uma} & \multirow{2}{*}{1} & 31.70 & -31.50 & $28.8 \pm 5.8$ & $0.3$ & \multirow{2}{*}{10.0} & \multirow{2}{*}{0.3} & \multirow{2}{*}{171.3} & \multirow{2}{*}{23.8} & \multirow{2}{*}{$29.9^{+13.5}_{-5.3}$} & \multirow{2}{*}{0.717} & \multirow{2}{*}{\citet{Shipp:2018}} \\
	 &  & 40.60 & -38.30 & $28.8 \pm 5.8$ & $0.3$ &  &  &  &  &  &  &  \\
	\bottomrule
\end{tabular}

\caption{The list of streams analyzed in this work (abridged version, see Table \ref{tab:streamprops-full} for full list).
For each stream, we compile two anchor points (usually the two endpoints of the stream) and provide their position on the sky, heliocentric distance, and angular uncertainty $\Delta\theta$.
We also calculate the stream length (on-sky angular separation between the two anchor points) and provide the stream width (Gaussian $\sigma$ perpendicular to the stream) reported in the reference.
The stream normal is reported in spherical coordinates $(l_\text{MW},b_\text{MW})$ that are aligned with the heliocentric Galactic system and centered on the Milky Way center.
The angular distance of the stream normal from the VPOS normal is $\theta_\text{obs}$ and the fraction of Monte Carlo stream normals that fall within the 10\% circle centered on the VPOS normal is $p_\text{inVPOS}$.}
\label{tab:streamprops}
\end{table*}

We apply the method described above to 64 Milky Way streams that extend more than 4$^\circ$ along the sky.
Our list of streams is based on the catalog in the \texttt{galstreams}\footnote{\href{https://github.com/cmateu/galstreams}{https://github.com/cmateu/galstreams}} package \citep{Mateu:2018} and augmented with more recent discoveries.
We exclude shorter streams (Ophiucus; \citealt{Bernard:2014}) and tidal deformations around some globular clusters (Eridanus and Pal 15; \citealt{Myeong:2017}).
We exclude cloud-like structures for which it is difficult to assign anchor points: EriPhe \citep{Li:2016}, Hercules-Aquila, TriAnd1, TriAnd2, and the Virgo and Pisces Overdensities \citep{GrillmairCarlin:2016}.
We also exclude the Monoceros Ring due to its disputed nature as a stream or a feature of the Milky Way disk \citep{Sheffield:2018, Laporte:2019}.
Additionally, we are forced to exclude the four streams (WG-1, WG-2, WG-3, WG-4) discovered in \citet{Agnello:2017} as they lack a distance measurement.
The stream data are compiled in Table \ref{tab:streamprops}, along with respective references.

With no established convention for naming newly discovered streams, a ``delightful anarchy'' has ensued \citep{GrillmairCarlin:2016}.
Unfortunately, a similar anarchy has pervaded in how {\it properties of the streams} have been reported.
Definitions of a stream footprint can include its endpoints, orbital pole, midpoint along the stream, or simply the coordinates of individual stars believed to be stream members.
The width of the streams, fit along a coordinate perpendicular to the stream, can be reported as the standard deviation of a Gaussian or as a full-width at half-maximum (FWHM) of the stellar density.
Further complications arise when studies provide these quantities in Galactocentric coordinates, sometimes without stating the assumptions required to transform from observed heliocentric coordinates.

In an attempt to wrangle this situation, we provide stream anchor points in a consistent format.
For each stream, we provide heliocentric on-sky positions of the anchor points, heliocentric distances to the anchors, and the Gaussian width of the stream perpendicular to the stream track.
If a stream width was reported as a FWHM in the reference, we convert this to a Gaussian standard deviation assuming the FWHM corresponds to that of a Gaussian ($\text{FWHM}=2.355\sigma$).

When compiling this catalog, three classes of references emerged.
Streams classified as Class 1 have references that report each of these quantities explicitly, allowing us to simply copy the values into our catalog with no extra work.
Typically we either adopt the width of the stream or the precision to which the endpoint is reported for the angular uncertainty on the endpoint $\Delta\theta$, whichever is the larger of the two.

Streams classified as Class 2 have references that provide distance measurements and widths explicitly, but do not explicitly provide endpoints.
In this case, we use the endpoints computed using \texttt{galstreams} and confirm that these match with figures of the stream in the reference.
As these endpoints are derived from representations of the stream, we assume an uncertainty on the endpoint position of $\Delta\theta = 0.5^\circ$ regardless of stream width.
Out of the 64 streams analyzed, 31 fall under Class 1 and 14 fall under Class 2.

Eighteen of the remaining streams required additional analysis to obtain anchor points.
For many of these, the endpoints computed by \texttt{galstreams} did not match the ends of the \texttt{galstreams} footprint and needed to be identified separately.
Others use overdensities along the stream (e.g.~the PAndAS stream) or globular clusters the stream originates from (e.g.~the NGC 5466 stream) rather than a poorly-resolved endpoint for the anchor.
Each of these cases is described in detail in Appendix \ref{app:redstreams}.

Finally, the Sagittarius stream has multiple wraps around the Milky Way and is not well described by two anchor points, but its orbital plane has been studied in detail \citep{Majewski:2003, Fellhauer:2006}.
We simply adopt the normal reported by \citet{Law:2010} of $(l_\text{MW}, b_\text{MW}) = (273.8, -14.5)$.
We include the Sagittarius stream in the results quoted below and assume it has $p_\text{inVPOS} = 0$.

\subsection{Results and comparison to isotropy} \label{subsec:streamresults}

\begin{figure}
\includegraphics[width=\linewidth]{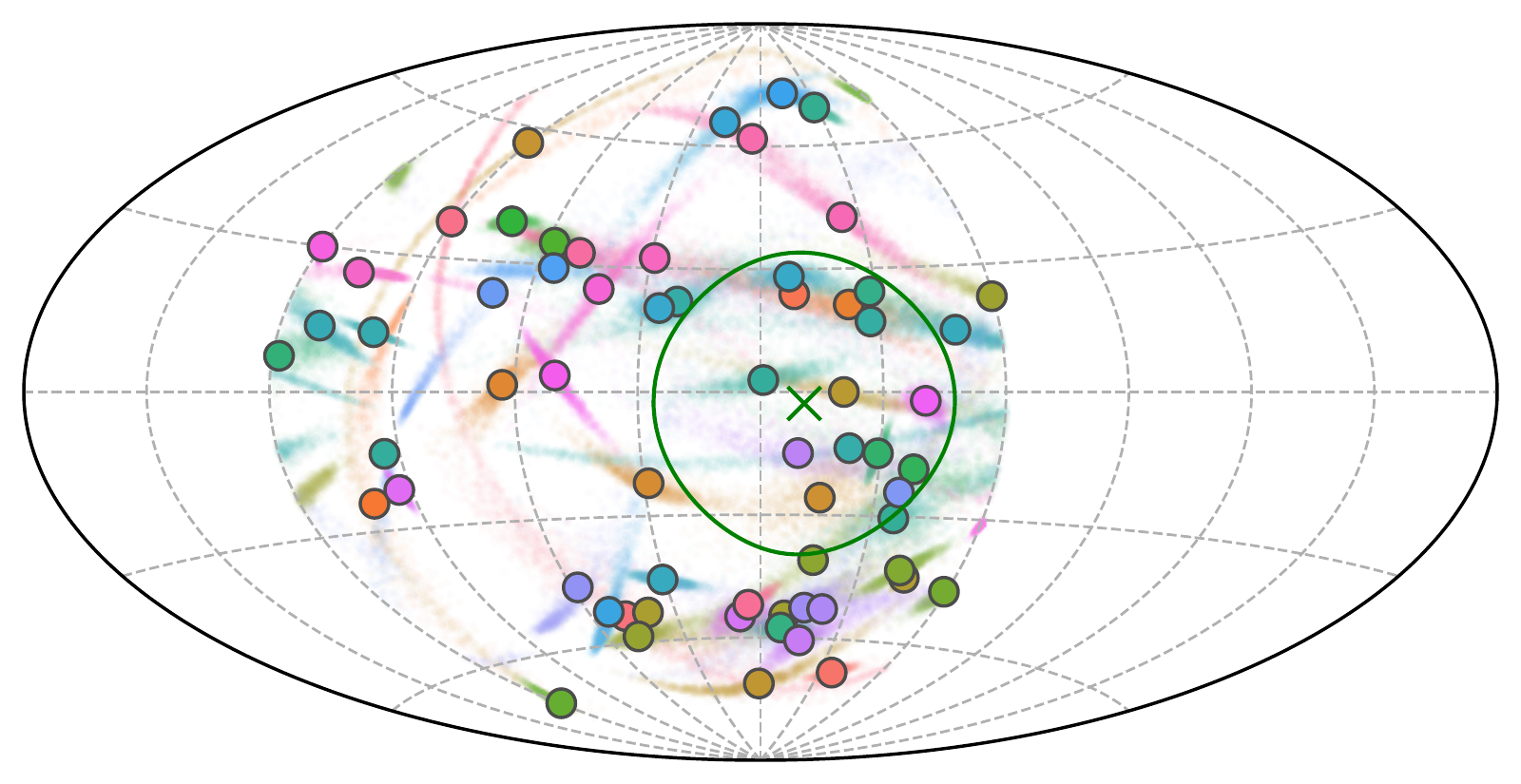}
\caption{All-sky distribution of normals for Milky Way streams.
Only one of the two normals (those from $120^\circ < l_\text{MW} < 300^\circ$) are shown.
The large points correspond to the observed (assuming no measurement errors) normals, while the point clouds show the results of 2,000 Monte Carlo simulations incorporating measurement errors.
The green circle contains 10\% of the sky around the assumed VPOS normal (marked as an ``x'').
{\it The Milky Way streams do not appear to preferentially align with the VPOS normal.}}
\label{fig:streamnormals}
\end{figure}

Figure \ref{fig:streamnormals} shows the on-sky distribution of Milky Way stream normals.
The figure includes the direction of the assumed VPOS normal $(l_\text{MW}, b_\text{MW}) = (169.3^\circ, -2.8^\circ)$ and the angular tolerance $\theta_\text{inVPOS} = 36.87^\circ$ around that direction.
Although the normal vectors are two-directional, since we do not know the orbital sense only one direction is plotted.
To match the range used in \citet{Pawlowski:2012} for easy comparison, only normals within $120^\circ < l_\text{MW} < 300^\circ$ are plotted.

For each stream, we calculate the angular distance $\theta_\text{obs}$ of the Monte Carlo stream normals from the VPOS normal and the fraction $p_\text{inVPOS}$ of Monte Carlo normals that fall within $\theta_\text{inVPOS}$ of the VPOS normal.
These quantities are presented in Table \ref{tab:streamprops}.

Overall, we do not find a tendency for stream normals to be clustered in the direction of the VPOS normal.
Of the 64 streams, 39 are excluded from VPOS membership ($p_\text{inVPOS} < 0.05$) and only 12 are likely VPOS members ($p_\text{inVPOS} > 0.5$).
The sum of membership probabilities for all streams is $\sum p_\text{inVPOS} = 13.32$.
The probability that at least $k=12$ out of $n=64$ uniformly distributed normals fall within the VPOS tolerance is 64.8\% (for $k=14$ this lowers to 40.2\%).

\begin{figure}
\includegraphics[width=\linewidth]{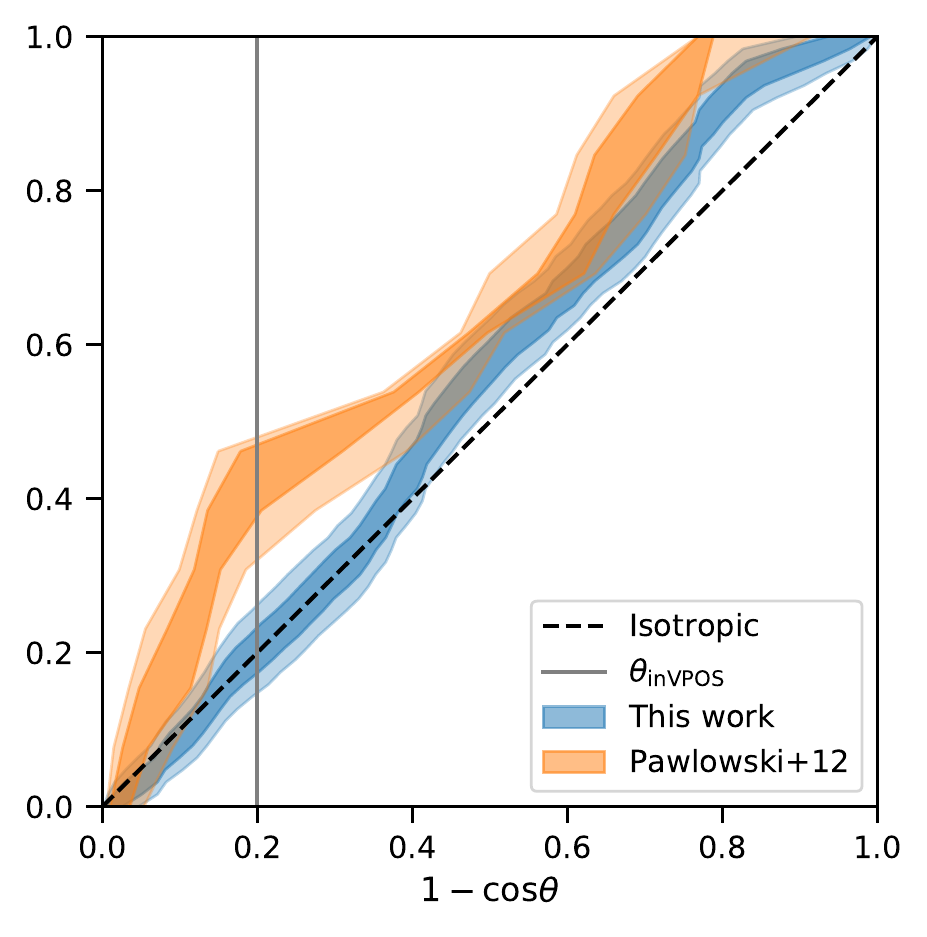}
\caption{Cumulative distribution of stream normals within a fraction of half-sphere measured from the VPOS normal.
We show the distribution of normals for streams used in this work (blue), the distribution of normals for the data compiled in \citet{Pawlowski:2012} (orange), the expectation for an isotropic distribution (black dashed), and the angular tolerance from the VPOS normal to be considered a member (grey solid).
Shaded bands corresond to the 1- and 2-$\sigma$ confidence intervals from Monte Carlo simulations.
{\it Our current census of stellar streams indicates no preferential alignment with the VPOS normal.}}
\label{fig:stream-cdf}
\end{figure}

The cumulative distribution of angular distances $\theta_\text{obs}$ between stream normals and the VPOS normal is plotted in Figure \ref{fig:stream-cdf}.
The shaded bands correspond to the 1- and 2-$\sigma$ confidence intervals as a result of the Monte Carlo simulations.
We also show results using the stream catalog from \citet{Pawlowski:2012} for comparison.
Adding in new stream discoveries has alleviated the difference between the observed stream normal distribution and an isotropic distribution (black dashed line).

Finally, it is possible that our position as observers leads us to over-represent streams that are not aligned with the VPOS.
Due to geometric effects, a stream defined by anchor points with heliocentric distance $d_\odot$ smaller than the distance between the Sun and the Galactic center $d_\text{GC}$ cannot lead to stream normals closer to $(l_\text{MW}, b_\text{MW}) = (180,0)^\circ$ or $(0,0)^\circ$ than
\begin{equation}
    \tan \alpha = \sqrt{\left( \frac{d_\odot}{d_\text{GC}} \right)^2 - 1}.
\end{equation}
Since the VPOS normal is close to $(180,0)^\circ$, streams that are close to the Sun (and thus easier to observe) might bias the overall sample against VPOS membership \citep{Pawlowski:2012}.

However, our results remain substantively unchanged after taking this effect into account.
When excluding all streams in the sample that have an anchor point with $d_\odot = 8.122$~kpc (15 streams with $p_\text{inVPOS} \simeq 0$), 24 of the remaining 49 streams have $p_\text{inVPOS} < 0.05$, 12 have $p_\text{inVPOS} > 0.5$, and $\sum p_\text{inVPOS} = 13.30$.
The probability that at least $k=12$ out of $n=49$ uniformly distributed normals fall within the VPOS tolerance is 26.5\% (for $k=14$ this lowers to 9.7\%).
It does not appear that this projection effect (in conjunction with many newly discovered streams being nearby) can explain the shift in the inferred distribution of stream normals with respect to the VPOS.

\subsection{Comparison to prior studies}

While some streams discovered since 2012 have been analyzed in the context of the VPOS, it has largely been a patchwork of techniques that are only applied to streams whose discoveries were announced in those studies.
Here, we compare our results for the alignment of individual streams with the VPOS to those from other studies.

Our results and those from prior studies that explicitly focus on the VPOS \citep{Pawlowski:2012, Pawlowski:2015} agree very well.
We find that most streams already classified as VPOS members (ATLAS, Cocytos, the Magellanic Stream, Pal 5, and Styx) all have $p_\text{inVPOS} \gtrsim 0.9$.
The two exceptions are Lethe and the NGC 5053 stream, which have normals located near the edge of the boundary for VPOS consideration, resulting in $p_\text{inVPOS} = 0.48$ even though they been classified as VPOS members.
The streams already classified as non-members (ACS, Acheron, Cetus, GD-1, NGC 5466, Orphan, PAndAS, Sagittarius, Triangulum/Pisces) all have $p_\text{inVPOS} = 0$.
We excluded the GCN stream \citep{Jin:2010} due to its tentative nature, though it is worth noting that it has been classified as a VPOS member.

Some studies have attempted to use orbit modeling to associate discovered streams with the VPOS, constraining the orbit using distances and stream tracks along the sky.
Using this method, \citet{Grillmair:2017a} found that PS1-D, Sangarius, and Scamander align with the VPOS and \citet{Grillmair:2017b} found that Molonglo and Murrumbidgee align with the VPOS while Kwando and Orinoco do not.
Strangely, we find nearly the opposite results: PS1-D, Sangarius, and Scamander all have $p_\text{inVPOS} = 0$, while Kwando, Molonglo, Murrumbidgee, and Orinoco have $p_\text{inVPOS} = $ 0.54, 0.53, 0.16, and 0.27, respectively.
This discrepancy merits further investigation, but may possibly be due to the potential assumed for the calculation \citep{Allen:1991}.

\citet{Shipp:2018} performed a search for stellar streams in the Dark Energy Survey, recovering four previously known streams and discovering 11 new ones.
They employed a very similar Monte Carlo technique as presented here to calculate stream normals (their Figure 17) and concluded that their stream sample did not preferentially align with the VPOS, without commenting on any individual streams that could be members.
However, their analysis did not account for the unknown orbital sense (such that a stream normal could be further than 90$^\circ$ from the VPOS normal) and effectively assumed the VPOS to have zero thickness ($\theta_\text{inVPOS} \simeq 0^\circ$).
Using their stream endpoint data, we find that 6 of the 15 streams in the DES footprint have $p_\text{inVPOS} > 0.5$ (ATLAS, Aliqa Uma, Elqui, Indus, Molonglo, and Ravi) and that $\sum p_\text{inVPOS} = 5.61$ for the entire sample.

Finally, some studies have analyzed cloud-like structures in the context of the VPOS.
\citet{Li:2016} discovered the EriPhe overdensity and noted that EriPhe, Hercules-Aquila, and the Virgo Overdensity fall along a polar great circle nearly aligned with the VPOS, suggesting that all three structures are remnants left by a single dwarf galaxy that was a VPOS member.
\citet{Boubert:2019} used {\it Gaia} RR Lyrae to characterize the Virgo Overdensity and noted the overdensity as aligning with the Magellanic Stream and the VPOS.
We excluded these cloud-like structures from our analysis because it is difficult to assign anchor points to such large, complex structures.

\subsection{Other clustering of stream normals}

While there is no clustering of stream normals in the VPOS direction, there does appear to be a cluster of normals in Figure \ref{fig:streamnormals} near $(l_\text{MW}, b_\text{MW}) = (180, -60)^\circ$.
Repeating the same calculations for this new direction confirms the clustering, with 11 streams (ACS, EBS, Fimbulthul, Fj\"{o}rm, Gaia-3, Kshir, Orphan, PS1-D, Sangarius, Scamander, Ylgr) having $p_\text{in-cluster} \simeq 1$ and the total $\sum p_\text{in-cluster} = 19.75$.
The probability that at least $k=19$ out of $n=64$ uniformly distributed normals fall within the assumed tolerance is 4.2\%.
The orbital poles for the young halo globular clusters are not concentrated in this direction.

There are several caveats to keep in mind regarding this apparent clustering.
Some of the streams have been claimed as a features/warps in the Milky Way's stellar disk (ACS, EBS) rather than the remnants of an accretion event \citep{Deason:2018, Laporte:2019}.
Several others are thought to be associated with each other \citep[GD-1 and Kshir;][]{Malhan:2019} or overlap in observed space such that an association is possible \citep[Orphan, PS1-D, Sangarius, and Scamander;][]{Grillmair:2017a}.
Removing even some of these streams from the catalog would alleviate the discrepancy from an isotropic distribution.
Finally, it is difficult to quantify the significance of this clustering without a handle on the selection function for detecting stellar streams and how that selection function projects onto orbital plane space.
Future spectroscopic confirmation of stream candidates and in-depth studies of this selection function will help to determine if this clustering of stream normals is significant.

\section{Orbit Modeling} \label{sec:orbits}

While we have shown that Milky Way streams and globular clusters as {\it systems} are not associated with the plane of satellites, it can still be interesting to associate {\it individual} objects that are VPOS members to attempt to establish a common origin (e.g.~group infall).
Independently of the VPOS, associating streams with known objects can point towards the progenitor of the stream and trace the process of hierarchical formation.
In this section, we integrate orbits for all of the Milky Way satellites (dwarf galaxies and globular clusters) to attempt to associate them with observed stellar streams.

\subsection{Methodology}

The full modelling of a satellite being unraveled into a stellar stream is challenging, as the results depend on the shape of the Galactic potential, internal and orbital kinematics of the progenitor, and the effects of dynamical friction and time-dependent substructures like the Galactic bar and the LMC.
For simplicity, and due to the fact that we have limited information of each stream (the two anchor points), we integrate orbits in static potentials that do not include the effects Galactic substructures or the LMC.

To associate a satellite with a stream, we integrate the satellite's orbit backwards in time for 1 Gyr using \texttt{gala} \citep{gala}.
For each object and stream pairing, we check if the orbit passes through the on-sky position of the stream's anchor points at approximately the right heliocentric distance.
If an orbit passes through both stream anchor points, within the corresponding 1-$\sigma$ angular and distance uncertainties (see Table \ref{tab:streamprops}), then it is considered an association.

Our initial conditions for the orbits of the globular clusters are the same as in Section \ref{sec:globulars}.
For the satellite galaxies, we use the compilation of position and velocity observations in Table A1 of \citet{Riley:2019}.\footnote{We adopt the proper motion measurements from \citet{Fritz:2018} except for the Magellanic Clouds, for which we use measurements from \citet{GaiaHelmi:2018}.}
We perform 2,000 Monte Carlo simulations of these initial conditions by sampling over uncertainties in the distance, radial velocity, and proper motions of each satellite.
By counting the fraction of Monte Carlo orbits for each satellite-stream pairing that are ``associated,'' we get a sense of which pairings are possible given the measurement uncertainties.

We repeat this procedure for three different Milky Way potentials that are implemented in \texttt{gala}: the Milky Way mass model in \citet{gala}, \texttt{MWPotential2014} as described in \citet{galpy}, and the potential used by \citet{Law:2010}.
Each of these potentials uses some combination of component potentials used to represent the dark matter halo and (static) Galactic baryonic structures.
In practice, these potentials vary the total mass and mass distributions while still being reasonably consistent with the latest constraints, allowing us to probe the effects of changing the potential on satellite-stream pairings.

\subsection{Results and the NGC 3201 -- \gjoll{} connection}

\begin{table}
\centering
\begin{tabular}{lccc}
	\toprule
	& $\omega$ Cen & NGC 3201 & NGC 4590 \\
	& Fimbulthul & \gjoll & PS1-E \\
	\midrule
	\citet{gala} & 0.000 & 0.577 & 0.012 \\
	\citet{galpy} & 0.189 & 0.271 & 0.141 \\
	\citet{Law:2010} & 0.035 & 0.398 & 0.000 \\
	\bottomrule
\end{tabular}
\caption{Potential associations between Milky Way satellites and known stellar streams based on orbit modeling.
Results are presented as the fraction of Monte Carlo orbit integrations that pass through both stream endpoints (see Section \ref{sec:orbits} for details).
Orbits are integrated in three different potentials \citep{gala, galpy, Law:2010}.
We recover the previously known association between $\omega$ Cen and Fimbulthul \citep{Ibata:2019Nature}, as well as a new possible association between NGC~3201 and \gjoll.}
\label{tab:goodpairings}
\end{table}

The main results of this procedure are shown in Table \ref{tab:goodpairings}.
We find only three satellite-stream pairings that were associated in 5\% or more of the 2,000 Monte Carlo simulations for {\it any} of the 3 potentials.
One of these is the previously reported association between $\omega$ Cen (NGC 5139) and the Fimbulthul stream \citep{Ibata:2019Nature}.
Of the three possible associations in Table \ref{tab:goodpairings}, all have $p_\text{inVPOS} = 0$.

Independently of the VPOS, we want to draw attention to the possible association between NGC 3201 and the \gjoll{} stellar stream.
This pairing had a high fraction of Monte Carlo orbit integrations pass through the stream relative to other pairings, independent of which Milky Way potential was adopted.
The cluster and the stream are also very close to each other; one only needs to integrate NGC 3201's orbit backwards $\sim$25 Myr to overlap with the stream.

NGC 3201 is a young halo cluster \citep{Mackey:2005} with a retrograde and eccentric orbit \citep{GaiaHelmi:2018}. 
The best-fit orbit of the \gjoll{} stream from \citet{Ibata:2019} lies in a similar parameter space as NGC 3201 in $E-L_z$ (our orbit integration results can be considered a natural manifestation of this fact).
Noting this similarity, \citet{Bianchini:2019} used {\it Gaia} DR2 data to identify tidal tails coming off of the cluster.
Taken together, these results suggest that NGC 3201 experienced tidal stripping that formed into the \gjoll{} stream.
Combining {\it Gaia} DR2 kinematics with independent elemental abundances to chemo-dynamically tag \gjoll{} stars as NGC 3201 members would further link these two structures (Hansen~et~al.~2020, in preparation).

\section{Discussion} \label{sec:discussion}

We have shown that the distributions of Milky Way globular cluster orbital poles (Section \ref{sec:globulars}) and stellar stream normals (Section \ref{sec:streams}) are not significantly clustered in the direction of the VPOS normal.
It can be tempting to use these results to reject the notion of a plane of satellites problem altogether.
However, we strongly caution against this.

The classical satellite galaxies of the Milky Way have been shown to have a highly-clustered distribution of orbital poles \citep{Pawlowski:2013}, a trend that has become more significant as higher-precision proper motion measurements become available \citep{Pawlowski:2020} and that extends to Milky Way ultrafaint satellites \citep{Fritz:2018}.
This observed clustering of satellite orbital poles is not consistent with expectations from cosmological simulations, regardless of whether the simulations considered are dark-matter-only \citep{Pawlowski:2014ELVIS, Pawlowski:2014, Ibata:2014} or include baryonic effects \citep{Muller:2018, Pawlowski:2019PHAT}.
There is also considerable evidence for rotationally-supported planes in other satellite systems within the Local Group, namely around M31 \citep{Ibata:2013, Santos-Santos:2019} and Centaurus A \citep{Muller:2018}. Taking this observed plane of satellite galaxies as given, we have to consider why other accreted material---young halo globular clusters and stellar streams---{\it doesn't} exhibit the same behavior.

The first possibility is that the plane of satellite galaxies represents a particularly stable orbital configuration and that the accreted clusters and streams are from progenitors that happened on easily consumed orbital trajectories.
The underlying physical reason could be related to the host (e.g.~the orientation of the disk), intrinsic properties of the progenitors (e.g.~less dense), or the orbits of the progenitors (e.g.~closer pericenters) that have somehow resulted in polar orbits being longer-lived.
This explanation can be tempting, as Sagittarius is in the process of disrupting and its orbital pole is located the furthest from the VPOS normal of any classical satellite \citep{Pawlowski:2020}.
However, this mechanism would lead to a {\it dearth} of orbital poles in the direction of the VPOS normal, rather than the observed distribution that is reasonably consistent with isotropy (see Figure \ref{fig:stream-cdf}).
Furthermore, satellite planes initialized in controlled simulations with static potentials disintegrate fairly rapidly (within 1-4~Gyr) due to subhalo-subhalo interactions, any asphericity of the potential, and deviations from a polar orientation \citep{Fernando:2017, Fernando:2018}.

The second possibility is that there is no ``safe'' orbital configuration and that the Milky Way's (stable) plane of satellites has formed recently, likely as a result of group infall.
This formulation agrees with the observed, nearly isotropic distribution of accreted material and opens the door to attributing many VPOS members to, for example, the LMC coming in on its first infall \citep{Besla:2007}.
However, using inferred infall times from comparing {\it Gaia} kinematics to subhalos in cosmological simulations \citep{Fillingham:2019}, high-probability VPOS satellite galaxy members have a range of infall times approximately 7-10 Gyr ago.
This is much longer than the 1-4 Gyr lifespan of satellite planes in controlled simulations, making the scenario of a recent group infall unlikely.
Additionally, detailed orbit integrations of 20 satellite galaxies (most of which are VPOS members) in a joint Milky Way, LMC, and SMC potential find that only half can be attributed to the LMC/SMC system \citep{Patel:2020}.

We also re-iterate that the orbital planes of globular clusters and streams effectively precess and nutate with time \citep{BinneyTremaine:2008, Erkal:2016}.
While modeling this process is subject to assumptions about the progenitor's orbit and properties of the host halo, including its total mass and triaxiality, \citet{Erkal:2016} quantified the orbital evolution of streams in aspherical potentials and found the precession rate for polar orbits to be $\sim$\.10$^\circ$~yr$^{-1}$.
Importantly for our results, they suggest that orbital planes aligned with the polar axis are the shortest-lived.
More detailed simulations of stream evolution, including examples in triaxial potentials, are warranted to see if the observed distribution of stream normals is consistent with theoretical expectations.

The association of stellar streams with the VPOS is further complicated by the gravitational presence of the LMC.
The LMC is a massive satellite that has been shown to perturb streams in the Southern Hemisphere, displacing a stream's systemic proper motion from its track on the sky \citep{Erkal:2019, Shipp:2019}.
Depending on the size of the offset (which has been observed to vary based on sky position, distance from the LMC, and stream width), this effect can make the stream normal computed using the two-anchor-point method a less reliable indicator of the orbital plane.
Detailed modeling of this effect in the context of the VPOS will require radial velocities for member stars of the affected streams, something that may be possible with upcoming data releases from the S$^5$ survey \citep{Li:2019}.

Finally, it is worth commenting on NGC 4147's relation to the Sagittarius dwarf galaxy.
NGC 4147 is classified as a Sagittarius globular cluster by \citet{Mackey:2005}, however it is currently counter-orbiting with respect to Sagittarius (see Figure \ref{fig:GCpoles}, lower right panel) while the other 5 Sagittarius clusters (NGC 6715, Terzan 7, Arp 2, Terzan 8, and Pal 12) have orbital poles that clump near Sagittarius's pole \citep{Vasiliev:2019}.
This confirms prior results based on HST proper motions \citep{Sohn:2018} and agrees with spectroscopic abundance measurements \citep{Villanova:2016} that suggest that the cluster is not associated with Sagittarius, but rather could be from {\it Gaia}-Enceladus \citep{Massari:2019}.

\section{Summary} \label{sec:summary}

In this work we have analyzed Milky Way globular clusters and stellar streams for dynamical alignment with the Milky Way's plane of satellites.
After computing globular cluster orbital poles and stream normals, we determine VPOS membership for each cluster and stream using Monte Carlo simulations to take observational uncertainties into account.
We compare the resulting distributions of orbital poles and stream normals to those expected from an isotropic distribution.
Finally, we use an orbit modeling scheme to attempt to associate Milky Way satellites with known stellar streams.
A summary of our main results is as follows:
\begin{itemize}
    \item Utilizing the latest globular cluster proper motions inferred from {\it Gaia} DR2 data, we find that globular cluster orbital poles are not clustered in the VPOS direction.
    The young halo clusters, the population expected to be associated with the VPOS, have the highest fraction of VPOS membership ($\sim$30\%), but this is substantially lower than the fraction of satellite galaxies associated with the VPOS.
    If the orbital poles were drawn from a uniform distribution, such an alignment around the VPOS would occur in at most 12.9\% of cases, much higher than the $\sim$0.1\% typically quoted in VPOS studies \citep{Pawlowski:2012}.
    \item Using our compilation of stream anchor points in Table \ref{tab:streamprops}, we similarly find that stellar stream normals are not clustered around the VPOS normal.
    Of the 64 streams analyzed here, 12 of them align with the VPOS; drawing from a uniform distribution results in similar clustering up to 64.8\% of the time.
    Streams discovered since prior analysis in \citet{Pawlowski:2012} have shifted the observed distribution of stream normals away from being clustered in the VPOS direction (Figure \ref{fig:stream-cdf}).
    These results are not substantially impacted by geometric effects due to our position as observers.
    \item Through our orbit modeling procedure, we find three possible instances of Milky Way globular clusters being associated with stellar streams.
    None of these pairings are of members of the VPOS, making orbit modeling an unlikely avenue for studying the VPOS as a consequence of hierarchical structure formation.
    \item Based on our orbit modeling, it quite likely that globular cluster NGC 3201 is connected to the \gjoll{} stream.
    Information from chemical abundances and radial velocity measurements of stream stars could confirm this association (Hansen~et~al.~2020, in preparation).
    \item It is very unlikely that NGC 4147 is a Sagittarius globular cluster, based on the fact that it is nearly counter-orbiting with respect to Sagittarius (Figure \ref{fig:GCpoles}, lower right panel).
\end{itemize}

Finally, we echo our caution against using our results to reject the notion of the plane of satellites problem.
The satellite galaxies of the Milky Way have been shown to have a highly-clustered distribution of orbital poles \citep{Fritz:2018, Pawlowski:2020}, an observation that is not consistent with expectations from $\Lambda$CDM \citep{Pawlowski:2019PHAT}.
However, our results do counter the notion that the Milky Way's plane of satellites is part of a pronounced Vast Polar Structure comprised of satellite galaxies, globular clusters, and stellar streams.
Our hope is that these results raise interesting questions about the stability of planar configurations and may guide future studies on satellite planes in the context of hierarchical structure formation.

\section*{Acknowledgements}

We thank Peter Ferguson, Davide Massari, Andrew Pace, and Marcel Pawlowski for their helpful comments, as well as an anonymous referee for their thoughtful reading.
AHR acknowledges support from a Texas A\&M University Merit Fellowship and an NSF Graduate Research Fellowship through Grant DGE-1746932.
LES acknowledges support from DOE Grant No. de-sc0010813.
This research made use of the Python Programming Language, along with many community-developed or maintained software packages including Astropy \citep{Astropy:13, Astropy:18}, gala \citep{gala, gala-zenodo}, Jupyter \citep{jupyter}, Matplotlib \citep{matplotlib}, NumPy \citep{numpy}, Pandas \citep{pandas}, SciPy \citep{scipy}, and Seaborn (\href{https://seaborn.pydata.org/}{seaborn.pydata.org}).
This research made extensive use of \href{https://arxiv.org/}{arXiv.org} and NASA's Astrophysics Data System for bibliographic information.

\bibliographystyle{mnras}
\bibliography{references,stream-refs,software}

\appendix

\section{Specific stream information} \label{app:redstreams}

Here we describe the choice of anchor points for streams that fall under Class 3 as described in Section \ref{subsec:streamcatalog}.

{\it 20.0-1}.
This stream was discovered by \citet{Mateu:2018} using RR Lyrae found in the Catalina Surveys \citep{Drake:2013a, Drake:2013b, Torrealba:2015}.
Its member stars nearly span across an entire great circle on the sky in 3 distinct strands.
The anchor points are chosen as the closest and furthest RR Lyrae members in heliocentric distance.
We take the reported stream width of 1.8 deg for the angular uncertainty and assume a 10\% error on the distances.

{\it Cetus}.
Originally noted by \citet{Yanny:2009} and confirmed by \citet{Newberg:2009}, the Cetus stellar stream was studied in detail by \citet{Yam:2013} using blue horizontal branch and red giant stars in SDSS DR8.
Its complicated structure is summarized in Table 1 of that paper, which characterizes the spatial and kinematic properties of the stream as a function of binned Galactic latitude; we derive our anchor points from the first and last rows of that table.
We use the midpoint of the Galactic latitude bin for the anchor point coordinate, the uncertainty in Galactic longitude for the angular uncertainty of the anchor, and the reported distance $d$ and uncertainty on that distance $\sigma_d$ for the distance and uncertainty of the anchor.

{\it Corvus}.
Corvus is the other high-confidence stream candidate discovered by \citet{Mateu:2018} using RR Lyrae from the Catalina Surveys.
Like the 20.0-1 stream, we use the closest and furthest RR Lyrae members in heliocentric distance for the anchor points.
We take the reported stream width of 1.63 deg for the angular uncertainty and assume a 10\% error on the distances.

{\it EBS}. Originally discovered by \citet{Grillmair:2006a}, the Eastern Banded Structure was re-examined in detail by \citet{Grillmair:2011} using the more complete coverage afforded by SDSS DR7.
The on-sky positions of the Northern and Southern ends of the stream are not reported explicitly, but can be identified using their Figure 1.
We assume an angular uncertainty of 0.5 deg for each endpoint.
Distances for each of these endpoints are reported explicitly, as well as the stream width.

{\it Hermus}.
We base our endpoints for the Hermus stellar stream on the \texttt{galstreams} footprint, which is based on the discovery paper \citep{Grillmair:2014}.
One endpoint is computed using the \texttt{galstreams} algorithm, while the other does not match with an end of the stream and is identified separately.
Distances for each of these endpoints are reported explicitly in \citet{Grillmair:2014}, as well as the stream width.

{\it Hyllus}.
We base our endpoints for the Hyllus stellar stream on the \texttt{galstreams} footprint, which is based on the discovery paper \citep{Grillmair:2014}.
Both endpoints computed using the \texttt{galstreams} algorithm do not match with the ends of the stream footprint and were identified separately.
Distances for each of these endpoints are reported explicitly in \citet{Grillmair:2014}, as well as the stream width.

{\it Kshir}.
Discovered using {\it Gaia} DR2 and confirmed spectroscopically by \citet{Malhan:2019}, the Kshir stream is distinct from but highly correlated in phase space with GD-1.
We use two spectroscopic member stars as anchor points, corresponding to the highest $\phi_1$ overall and the lowest $\phi_1$ star from SDSS.
These stars were chosen due to how close they are to the best-fit orbit of Kshir and their high separation from each other.
We assume both anchors are at the uncertainty-weighted distance of the stream (10 kpc) with $\pm1$~kpc uncertainty, while the angular uncertainties are both assumed to be 0.5 degrees.

{\it Kwando}.
We base our endpoints for the Kwando stellar stream on the \texttt{galstreams} footprint, which is based on the discovery paper \citep{Grillmair:2014}.
Both endpoints computed using the \texttt{galstreams} algorithm do not match with the ends of the stream footprint and were identified separately.
\citet{Grillmair:2017b} reports the FWHM of the stream as 22 arcmin (Gaussian $\sigma = 0.16$ deg) and a physical width of 130 pc, corresponding to a heliocentric distance of 20.3 kpc.
We assume both anchor points are at this distance and that the distance uncertainty is 20\%.

{\it Magellanic Stream}.
While the Magellanic Stream has been studied with great detail in recent years \citep{Nidever:2008, Nidever:2010}, its sheer size and complexity does not easily lend to being described by anchor points.
We opt for the anchor points adopted by \citet{Pawlowski:2012}, which correspond to overdensities identified by \citet{Bruns:2005}.
We adopt the same 55~kpc distance for both anchor points, with a 20\% distance uncertainty and an angular uncertainty of 2 degrees.

{\it Murrumbidgee}.
We base our endpoints for the Murrumbidgee stellar stream on the \texttt{galstreams} footprint, which is based on the discovery paper \citep{Grillmair:2014}.
Both endpoints computed using the \texttt{galstreams} algorithm do not match with the ends of the stream footprint and were identified separately.
\citet{Grillmair:2017b} quotes a heliocentric distance of 20 kpc.
We assume both anchor points are at this distance and that the distance uncertainty is 20\%.

{\it NGC 5053}. Using SDSS data, \citet{Lauchner:2006} found a stellar tidal debris stream extending in one direction from globular cluster NGC 5053.
We mimic \citet{Pawlowski:2012} and use the globular cluster as one anchor point (using spatial data from \citealt{Harris:1996}, 2010 edition) and the quoted end of the stream in \citet{Lauchner:2006} as the other.
The error in distance to the globular cluster is assumed to be a distance modulus error of 0.1 mag \citep{Vasiliev:2019}, while the stream endpoint is assumed to have 20\% uncertainty.
The angular uncertainty to the cluster is assumed to be zero, while the angular uncertainty to the stream endpoint is assumed to be 1 degree.

{\it NGC 5466}. There is a long stellar tidal tail associated with globular cluster NGC 5466 \citep{Grillmair:2006b}.
We mimic \citet{Pawlowski:2012} and use the globular cluster as one anchor point (using spatial data from \citealt{Harris:1996}, 2010 edition) and the quoted end of the stream in \citet{Grillmair:2006b} as the other.
The error in distance to the globular cluster is assumed to be a distance modulus error of 0.1 mag \citep{Vasiliev:2019}, while the stream endpoint is assumed to have 20\% uncertainty.
The angular uncertainty to the cluster is assumed to be zero, while the angular uncertainty to the stream endpoint is assumed to be 2 degrees.

{\it NGC 5904}. Using {\it Gaia} DR2 data, \citet{Grillmair:2019} detected a 50$^\circ$ long stream extending westward from NGC 5904 (M5) believed to be the trailing arm.
Similar to NGCs 5053 and 5466, we use the globular cluster as one anchor point (using spatial data from \citealt{Harris:1996}, 2010 edition) and the quoted western end of the stream for the other.
The error in distance to the globular cluster is assumed to be a distance modulus error of 0.1 mag \citep{Vasiliev:2019}, while the stream endpoint is assumed to have 20\% uncertainty.
The angular uncertainty to the cluster is assumed to be zero, while the angular uncertainty to the stream endpoint is assumed to be 1 degree.

{\it Orinoco}.
We base our endpoints for the Orinoco stellar stream on the \texttt{galstreams} footprint, which refers to \citet{Grillmair:2017b}.
Both endpoints computed using the \texttt{galstreams} algorithm do not match with the ends of the stream footprint and were identified separately.
\citet{Grillmair:2017b} quotes a heliocentric distance of $20\pm 3$ kpc.
We assume both anchor points are at this distance.

{\it Pal 5}. There is a long stellar tidal tail associated with the Pal 5 cluster \citep{Odenkirchen:2001, GrillmairDionatos:2006}. 
Recently, \citet{Starkman:2019} used Gaia DR2 data to reveal an additional 7 degrees of the leading arm.
We use the endpoints quoted in their Figure 4, transformed to (RA, Dec) using the coordinate transformation described in their Appendix B.
Both endpoints are assumed to be at the distance to the cluster of 23.2 kpc \citep[][2010 edition]{Harris:1996} with distance errors assumed to be 10\%.

{\it PAndAS}.
This stream was discovered as a Milky Way foreground structure in the Pan-Andromeda Archaeological Survey centered on M31 \citep{Martin:2014}.
We use the overdensity/possible progenitor located at $(l,b) = (117.2, −16.6)$ for one anchor point and the southern end of the stream as the other \citep{Deason:2014}.
Both anchor points are assumed to be at the average distance to the stream of $17\pm3$ kpc.

{\it Sangarius}. 
We base our endpoints for the Sangarius stellar stream on the \texttt{galstreams} footprint, which is based on the discovery paper \citep{Grillmair:2017a}.
Both endpoints computed using the \texttt{galstreams} algorithm do not match with the ends of the stream footprint and were identified separately.
We assume both endpoints are at $21\pm5$ kpc \citep{Grillmair:2017b} and that the stream FWHM is 1 deg FWHM (Gaussian $\sigma = 0.422$ deg).

{\it Scamander}.
We base our endpoints for the Scamander stellar stream on the \texttt{galstreams} footprint, which is based on the discovery paper \citep{Grillmair:2017a}.
Both endpoints computed using the \texttt{galstreams} algorithm do not match with the ends of the stream footprint and were identified separately.
We assume both endpoints are at $21\pm5$ kpc \citep{Grillmair:2017b} and that the stream FWHM is 1 deg FWHM (Gaussian $\sigma = 0.422$ deg).

\section{Full versions of tables} \label{app:fulltables}

\begin{table*}
\centering
\begingroup
\renewcommand{\arraystretch}{1.25}
\begin{tabular}{lcccccrccc}
	\toprule
	Name & Type & $l_\text{pole}$ & $b_\text{pole}$ & $\Delta_\text{pole}$ & $\theta_\text{pred}$ & $\theta_\text{obs}$ & $p_\text{inVPOS}$ & $p_\text{>VPOS}$ & $p_\text{>obs}$ \\
	 &  & [deg] & [deg] & [deg] & [deg] & [deg] &  &  &  \\
	\midrule
	NGC 104 & BD & 52.5 & -61.7 & 1.4 & 51.1 & $-80.2^{+0.1}_{-0.1}$ & 0.000 & 0.000 & 0.000 \\
	NGC 288 & OH & 133.2 & 32.6 & 8.1 & 44.7 & $49.2^{+4.2}_{-3.0}$ & 0.000 & 0.002 & 0.000 \\
	NGC 362 & YH & 143.7 & 9.6 & 15.7 & 27.0 & $29.1^{+2.2}_{-0.9}$ & 0.946 & 0.040 & 0.065 \\
	Whiting 1 & UN & 284.1 & -20.8 & 5.5 & 43.7 & $-67.9^{+3.8}_{-3.4}$ & 0.000 & 0.000 & 0.000 \\
	NGC 1261 & YH & 265.2 & 39.7 & 17.2 & 21.6 & $-79.4^{+7.2}_{-14.0}$ & 0.004 & 0.038 & 0.000 \\
	Pal 1 & BD & 183.2 & -75.7 & 1.2 & 66.9 & $73.4^{+0.4}_{-0.5}$ & 0.000 & 0.000 & 0.000 \\
	E 1 & YH & 216.9 & 36.4 & 41.0 & 6.5 & $56.9^{+20.7}_{-27.4}$ & 0.223 & 0.211 & 0.086 \\
	Eridanus & YH & 148.9 & 27.7 & 42.6 & 35.7 & $41.5^{+18.2}_{-5.4}$ & 0.240 & 0.282 & 0.288 \\
	Pal 2 & YH & 263.1 & -2.9 & 67.9 & 84.7 & $-86.2^{+2.5}_{-1.4}$ & 0.000 & 0.578 & 0.000 \\
	NGC 1851 & OH & 297.4 & 18.0 & 9.3 & 40.3 & $-53.0^{+5.9}_{-3.6}$ & 0.000 & 0.003 & 0.000 \\
	NGC 1904 & OH & 116.0 & -4.6 & 21.2 & 49.0 & $53.2^{+11.9}_{-4.2}$ & 0.000 & 0.079 & 0.021 \\
	NGC 2298 & OH & 134.7 & 34.2 & 4.7 & 41.7 & $49.2^{+1.6}_{-1.5}$ & 0.000 & 0.000 & 0.000 \\
	NGC 2419 & OH & 254.4 & -32.6 & 27.6 & 61.9 & $81.3^{+6.2}_{-9.1}$ & 0.000 & 0.142 & 0.003 \\
	Pyxis & YH & 158.9 & 9.5 & 3.7 & 8.8 & $15.9^{+3.2}_{-2.6}$ & 1.000 & 0.000 & 0.000 \\
	NGC 2808 & OH & 90.9 & -78.5 & 6.0 & 23.1 & $85.0^{+3.6}_{-5.9}$ & 0.000 & 0.000 & 0.000 \\
	E 3 & UN & 112.2 & -60.2 & 1.2 & 25.4 & $71.8^{+1.3}_{-1.1}$ & 0.000 & 0.000 & 0.000 \\
	Pal 3 & YH & 164.8 & -22.1 & 28.1 & 16.6 & $26.4^{+16.2}_{-8.8}$ & 0.753 & 0.167 & 0.463 \\
	NGC 3201 & YH & 113.2 & 62.6 & 0.5 & 46.1 & $77.7^{+0.5}_{-0.4}$ & 0.000 & 0.000 & 0.000 \\
	Pal 4 & YH & 64.5 & 15.4 & 74.8 & 16.2 & $-58.4^{+20.1}_{-29.6}$ & 0.225 & 0.362 & 0.000 \\
	Crater & UN & 156.3 & 20.1 & 75.7 & 9.4 & $49.7^{+26.3}_{-30.4}$ & 0.347 & 0.502 & 0.646 \\
	NGC 4147 & SG & 81.8 & 16.0 & 29.5 & 20.4 & $68.9^{+14.9}_{-17.5}$ & 0.028 & 0.231 & 0.000 \\
	NGC 4372 & OH & 119.9 & -59.5 & 1.6 & 35.9 & $68.2^{+1.6}_{-1.4}$ & 0.000 & 0.000 & 0.000 \\
	Rup 106 & YH & 201.5 & -44.1 & 3.0 & 19.0 & $50.2^{+2.7}_{-2.6}$ & 0.000 & 0.000 & 0.000 \\
	NGC 4590 & YH & 207.4 & -48.6 & 1.3 & 12.8 & $56.2^{+0.6}_{-0.6}$ & 0.000 & 0.000 & 0.000 \\
	NGC 4833 & OH & 329.5 & -51.1 & 8.1 & 29.0 & $-56.7^{+5.5}_{-5.8}$ & 0.000 & 0.000 & 0.000 \\
	NGC 5024 & OH & 166.0 & -15.0 & 2.8 & 12.6 & $12.9^{+1.0}_{-0.7}$ & 1.000 & 0.000 & 0.000 \\
	NGC 5053 & YH & 155.9 & -13.3 & 2.2 & 12.5 & $17.1^{+1.1}_{-1.1}$ & 1.000 & 0.000 & 0.000 \\
	NGC 5139 & UN & 323.3 & 50.4 & 2.2 & 39.1 & $-52.4^{+2.2}_{-2.1}$ & 0.000 & 0.000 & 0.000 \\
	NGC 5272 & YH & 189.0 & -32.3 & 2.9 & 31.2 & $34.7^{+2.3}_{-2.3}$ & 0.827 & 0.000 & 0.000 \\
	NGC 5286 & OH & 165.4 & 38.3 & 5.9 & 8.0 & $41.2^{+5.1}_{-5.9}$ & 0.227 & 0.000 & 0.000 \\
	NGC 5466 & YH & 300.0 & 18.0 & 2.3 & 15.8 & $-50.6^{+2.2}_{-1.9}$ & 0.000 & 0.000 & 0.000 \\
	NGC 5634 & OH & 296.6 & -22.5 & 13.8 & 26.1 & $-57.1^{+13.4}_{-9.4}$ & 0.006 & 0.014 & 0.000 \\
	NGC 5694 & OH & 92.6 & 41.2 & 18.0 & 47.3 & $79.4^{+7.6}_{-10.0}$ & 0.000 & 0.035 & 0.000 \\
	IC 4499 & YH & 358.3 & 24.1 & 2.3 & 17.7 & $-23.0^{+1.8}_{-1.9}$ & 1.000 & 0.000 & 0.000 \\
	NGC 5824 & OH & 33.3 & -32.9 & 3.1 & 54.7 & $-54.9^{+0.5}_{-0.6}$ & 0.000 & 0.000 & 0.000 \\
	Pal 5 & YH & 346.0 & -25.1 & 2.7 & 28.1 & $-28.1^{+1.0}_{-1.0}$ & 1.000 & 0.000 & 0.000 \\
	NGC 5897 & OH & 325.3 & -29.9 & 2.8 & 24.5 & $-40.1^{+2.1}_{-2.4}$ & 0.101 & 0.000 & 0.000 \\
	NGC 5904 & OH & 114.8 & -16.0 & 4.6 & 25.9 & $55.1^{+3.1}_{-2.9}$ & 0.000 & 0.000 & 0.000 \\
	NGC 5927 & BD & 228.4 & -81.3 & 0.7 & 10.9 & $82.8^{+0.5}_{-0.5}$ & 0.000 & 0.000 & 0.000 \\
	NGC 5946 & OH & 190.0 & -13.8 & 6.7 & 19.0 & $23.2^{+5.7}_{-6.4}$ & 0.993 & 0.000 & 0.000 \\
	BH 176 & BD & 298.1 & -83.5 & 1.4 & 49.5 & $-88.7^{+0.8}_{-0.9}$ & 0.000 & 0.000 & 0.000 \\
	NGC 5986 & OH & 0.9 & -25.5 & 9.7 & 25.2 & $-31.1^{+6.3}_{-8.4}$ & 0.822 & 0.004 & 0.000 \\
	FSR 1716 & UN & 6.5 & -59.9 & 2.5 & 11.5 & $-64.2^{+1.4}_{-1.3}$ & 0.000 & 0.000 & 0.000 \\
	Pal 14 & YH & 3.3 & -40.1 & 68.6 & 32.5 & $-48.4^{+24.7}_{-14.1}$ & 0.252 & 0.398 & 0.000 \\
	BH 184 & BD & 349.8 & -54.1 & 3.0 & 6.8 & $-57.1^{+0.9}_{-0.9}$ & 0.000 & 0.000 & 0.000 \\
	NGC 6093 & OH & 26.6 & -9.7 & 6.0 & 25.3 & $-39.4^{+2.2}_{-2.0}$ & 0.104 & 0.000 & 0.000 \\
	NGC 6121 & OH & 275.2 & 19.4 & 69.1 & 73.7 & $-77.6^{+8.1}_{-2.5}$ & 0.000 & 0.812 & 0.000 \\
	NGC 6101 & OH & 342.7 & 53.4 & 1.4 & 23.3 & $-50.9^{+0.5}_{-0.5}$ & 0.000 & 0.000 & 0.000 \\
	NGC 6144 & OH & 121.3 & 26.8 & 3.9 & 15.0 & $54.9^{+2.1}_{-2.0}$ & 0.000 & 0.000 & 0.000 \\
	NGC 6139 & OH & 217.1 & -29.4 & 6.2 & 34.7 & $52.5^{+5.2}_{-6.7}$ & 0.016 & 0.000 & 0.000 \\
	\bottomrule
\end{tabular}
\endgroup
\caption{VPOS membership statistics for all globular clusters analyzed in this work.
For column descriptions, see Table \ref{tab:GCprops}.}
\label{tab:GCprops-full}
\end{table*}

\begin{table*}
\centering
\begingroup
\renewcommand{\arraystretch}{1.25}
\begin{tabular}{lcccccrccc}
	\toprule
	Name & Type & $l_\text{pole}$ & $b_\text{pole}$ & $\Delta_\text{pole}$ & $\theta_\text{pred}$ & $\theta_\text{obs}$ & $p_\text{inVPOS}$ & $p_\text{>VPOS}$ & $p_\text{>obs}$ \\
	 &  & [deg] & [deg] & [deg] & [deg] & [deg] &  &  &  \\
	\midrule
	Terzan 3 & BD & 212.5 & -46.3 & 5.0 & 3.7 & $57.6^{+0.9}_{-0.6}$ & 0.000 & 0.000 & 0.000 \\
	NGC 6171 & OH & 208.7 & -35.6 & 4.7 & 39.2 & $48.9^{+4.0}_{-4.3}$ & 0.005 & 0.000 & 0.000 \\
	ESO 452-11 & YH & 14.7 & 12.0 & 20.7 & 5.9 & $-27.2^{+11.5}_{-15.3}$ & 0.791 & 0.072 & 0.000 \\
	NGC 6205 & OH & 243.4 & 20.3 & 3.0 & 44.0 & $76.1^{+1.4}_{-1.0}$ & 0.000 & 0.000 & 0.000 \\
	NGC 6229 & YH & 13.9 & -11.2 & 26.6 & 9.0 & $-29.6^{+19.6}_{-16.0}$ & 0.652 & 0.104 & 0.000 \\
	NGC 6218 & OH & 212.9 & -51.7 & 2.5 & 59.4 & $60.8^{+2.2}_{-2.3}$ & 0.000 & 0.000 & 0.000 \\
	FSR 1735 & UN & 207.3 & 1.9 & 5.3 & 37.4 & $38.4^{+4.2}_{-5.1}$ & 0.369 & 0.000 & 0.000 \\
	NGC 6235 & OH & 40.6 & -39.2 & 4.6 & 51.2 & $-63.2^{+1.2}_{-1.2}$ & 0.000 & 0.000 & 0.000 \\
	NGC 6254 & OH & 106.7 & -52.0 & 2.5 & 65.2 & $71.3^{+0.2}_{-0.1}$ & 0.000 & 0.000 & 0.000 \\
	NGC 6256 & BD & 223.9 & -11.0 & 6.8 & 51.6 & $54.6^{+4.0}_{-4.2}$ & 0.000 & 0.000 & 0.000 \\
	Pal 15 & OH & 108.5 & -8.6 & 37.9 & 48.5 & $60.0^{+18.5}_{-10.0}$ & 0.000 & 0.294 & 0.087 \\
	NGC 6266 & OH & 169.8 & -55.4 & 7.1 & 41.4 & $52.9^{+4.1}_{-3.9}$ & 0.000 & 0.000 & 0.000 \\
	NGC 6273 & OH & 167.2 & 24.1 & 13.2 & 25.8 & $26.7^{+9.3}_{-6.6}$ & 0.861 & 0.006 & 0.038 \\
	NGC 6284 & OH & 268.0 & -4.3 & 7.4 & 70.9 & $-81.6^{+2.5}_{-2.0}$ & 0.000 & 0.000 & 0.000 \\
	NGC 6287 & OH & 262.5 & 5.3 & 2.1 & 33.8 & $-86.4^{+2.0}_{-1.9}$ & 0.000 & 0.000 & 0.000 \\
	NGC 6293 & OH & 149.3 & 45.3 & 27.3 & 48.5 & $51.7^{+10.6}_{-10.2}$ & 0.000 & 0.142 & 0.070 \\
	NGC 6304 & BD & 139.8 & -69.0 & 2.6 & 63.1 & $69.1^{+2.1}_{-2.5}$ & 0.000 & 0.000 & 0.000 \\
	NGC 6316 & BD & 293.8 & -52.6 & 9.5 & 68.1 & $-72.4^{+1.6}_{-2.6}$ & 0.000 & 0.005 & 0.000 \\
	NGC 6341 & OH & 44.3 & -5.1 & 6.8 & 41.6 & $-55.4^{+5.6}_{-4.6}$ & 0.000 & 0.000 & 0.000 \\
	NGC 6325 & OH & 18.0 & 17.2 & 16.0 & 18.0 & $-31.8^{+13.4}_{-9.5}$ & 0.653 & 0.019 & 0.000 \\
	NGC 6333 & OH & 105.2 & -29.4 & 3.0 & 15.5 & $66.2^{+0.5}_{-0.4}$ & 0.000 & 0.000 & 0.000 \\
	NGC 6342 & BD & 79.7 & -27.4 & 3.8 & 5.7 & $87.8^{+1.6}_{-2.7}$ & 0.000 & 0.000 & 0.000 \\
	NGC 6356 & BD & 310.9 & -48.8 & 2.5 & 59.2 & $-61.2^{+1.4}_{-1.7}$ & 0.000 & 0.000 & 0.000 \\
	NGC 6355 & OH & 251.2 & 17.4 & 6.7 & 52.4 & $83.1^{+3.4}_{-5.2}$ & 0.000 & 0.002 & 0.000 \\
	NGC 6352 & BD & 51.5 & -77.7 & 1.0 & 47.0 & $-87.1^{+0.5}_{-0.5}$ & 0.000 & 0.000 & 0.000 \\
	IC 1257 & OH & 205.5 & 68.5 & 19.1 & 51.0 & $75.0^{+8.5}_{-7.7}$ & 0.000 & 0.047 & 0.003 \\
	Terzan 2 & BD & 31.1 & 69.3 & 19.4 & 37.9 & $-71.8^{+2.1}_{-3.8}$ & 0.000 & 0.044 & 0.000 \\
	NGC 6366 & OH & 241.0 & -56.2 & 2.0 & 76.1 & $77.6^{+0.9}_{-0.9}$ & 0.000 & 0.000 & 0.000 \\
	Terzan 4 & OH & 124.1 & -34.5 & 11.9 & 49.7 & $52.5^{+7.5}_{-11.7}$ & 0.106 & 0.018 & 0.003 \\
	BH 229 & OH & 16.4 & 7.9 & 37.2 & 17.8 & $-36.8^{+23.0}_{-22.0}$ & 0.502 & 0.359 & 0.000 \\
	NGC 6362 & OH & 124.7 & -43.1 & 3.3 & 16.5 & $56.5^{+3.3}_{-3.1}$ & 0.000 & 0.000 & 0.000 \\
	NGC 6380 & BD & 292.0 & 76.3 & 9.9 & 61.4 & $-79.5^{+4.6}_{-4.3}$ & 0.000 & 0.003 & 0.000 \\
	Terzan 1 & YH & 120.8 & -76.2 & 8.8 & 66.8 & $78.3^{+2.3}_{-4.4}$ & 0.000 & 0.008 & 0.001 \\
	Ton 2 & BD & 153.9 & -47.6 & 13.0 & 7.2 & $46.7^{+7.7}_{-6.3}$ & 0.000 & 0.013 & 0.003 \\
	NGC 6388 & BD & 342.9 & 60.5 & 4.1 & 35.4 & $-57.9^{+2.9}_{-3.8}$ & 0.000 & 0.000 & 0.000 \\
	NGC 6402 & OH & 36.4 & -42.1 & 4.8 & 3.3 & $-62.1^{+4.5}_{-3.1}$ & 0.000 & 0.000 & 0.000 \\
	NGC 6401 & OH & 262.4 & 52.6 & 1.7 & 61.7 & $-86.0^{+1.5}_{-1.2}$ & 0.000 & 0.000 & 0.000 \\
	NGC 6397 & OH & 282.2 & -44.7 & 2.5 & 71.4 & $-76.0^{+1.1}_{-1.0}$ & 0.000 & 0.000 & 0.000 \\
	Pal 6 & YH & 85.1 & -3.9 & 3.5 & 80.8 & $83.9^{+1.3}_{-1.0}$ & 0.000 & 0.000 & 0.000 \\
	NGC 6426 & YH & 14.6 & -63.3 & 3.8 & 32.5 & $-68.7^{+2.9}_{-3.2}$ & 0.000 & 0.000 & 0.000 \\
	Djorg 1 & UN & 125.4 & -69.9 & 5.2 & 65.1 & $72.9^{+1.8}_{-2.0}$ & 0.000 & 0.000 & 0.000 \\
	Terzan 5 & BD & 238.0 & -52.3 & 13.5 & 74.8 & $75.1^{+4.1}_{-6.8}$ & 0.000 & 0.025 & 0.001 \\
	NGC 6440 & BD & 184.5 & 33.5 & 16.6 & 4.2 & $39.5^{+7.2}_{-7.6}$ & 0.377 & 0.040 & 0.034 \\
	NGC 6441 & BD & 124.3 & -71.0 & 3.3 & 68.7 & $74.1^{+0.8}_{-1.0}$ & 0.000 & 0.000 & 0.000 \\
	Terzan 6 & BD & 133.4 & 72.3 & 11.4 & 70.5 & $78.1^{+1.4}_{-1.8}$ & 0.000 & 0.007 & 0.000 \\
	NGC 6453 & OH & 79.1 & -14.5 & 3.2 & 74.2 & $88.6^{+0.9}_{-1.3}$ & 0.000 & 0.000 & 0.000 \\
	NGC 6496 & BD & 116.1 & -59.0 & 1.7 & 47.8 & $69.6^{+0.5}_{-0.5}$ & 0.000 & 0.000 & 0.000 \\
	Terzan 9 & YH & 249.8 & -13.7 & 10.1 & 74.5 & $79.8^{+6.0}_{-9.2}$ & 0.000 & 0.005 & 0.000 \\
	Djorg 2 & BD & 306.1 & -78.5 & 2.0 & 84.4 & $-84.4^{+1.1}_{-1.7}$ & 0.000 & 0.000 & 0.000 \\
	NGC 6517 & OH & 138.6 & -36.7 & 5.9 & 17.1 & $44.0^{+5.7}_{-5.5}$ & 0.102 & 0.000 & 0.000 \\
	\bottomrule
\end{tabular}
\endgroup
\caption{Continuation of Table \ref{tab:GCprops-full}.}
\end{table*}

\begin{table*}
\centering
\begingroup
\renewcommand{\arraystretch}{1.25}
\begin{tabular}{lcccccrccc}
	\toprule
	Name & Type & $l_\text{pole}$ & $b_\text{pole}$ & $\Delta_\text{pole}$ & $\theta_\text{pred}$ & $\theta_\text{obs}$ & $p_\text{inVPOS}$ & $p_\text{>VPOS}$ & $p_\text{>obs}$ \\
	 &  & [deg] & [deg] & [deg] & [deg] & [deg] &  &  &  \\
	\midrule
	Terzan 10 & BD & 286.9 & -18.6 & 3.1 & 57.8 & $-65.0^{+1.9}_{-2.7}$ & 0.000 & 0.000 & 0.000 \\
	NGC 6522 & OH & 284.4 & -24.8 & 48.6 & 43.8 & $-70.1^{+4.1}_{-12.5}$ & 0.061 & 0.094 & 0.000 \\
	NGC 6535 & OH & 338.1 & 68.6 & 2.6 & 41.9 & $-66.1^{+2.3}_{-2.4}$ & 0.000 & 0.000 & 0.000 \\
	NGC 6528 & BD & 274.2 & -16.9 & 4.2 & 27.6 & $-76.6^{+2.0}_{-1.3}$ & 0.000 & 0.000 & 0.000 \\
	NGC 6539 & BD & 29.9 & -33.4 & 5.8 & 26.0 & $-52.5^{+5.5}_{-4.9}$ & 0.000 & 0.000 & 0.000 \\
	NGC 6540 & OH & 70.1 & -66.4 & 2.0 & 84.4 & $-88.9^{+0.7}_{-1.0}$ & 0.000 & 0.000 & 0.000 \\
	NGC 6544 & OH & 266.5 & 2.9 & 7.1 & 82.5 & $-82.7^{+0.2}_{-0.1}$ & 0.000 & 0.000 & 0.000 \\
	NGC 6541 & OH & 53.7 & -48.3 & 5.5 & 18.6 & $-75.7^{+2.6}_{-2.3}$ & 0.000 & 0.000 & 0.000 \\
	ESO 280-06 & UN & 76.2 & -19.9 & 11.3 & 65.3 & $-86.7^{+2.4}_{-4.4}$ & 0.000 & 0.004 & 0.000 \\
	NGC 6553 & BD & 349.9 & -82.0 & 1.5 & 83.7 & $-84.8^{+1.3}_{-1.5}$ & 0.000 & 0.000 & 0.000 \\
	NGC 6558 & OH & 66.2 & -20.3 & 10.1 & 46.8 & $-78.9^{+0.8}_{-0.7}$ & 0.000 & 0.000 & 0.000 \\
	Pal 7 & BD & 102.5 & -78.4 & 0.5 & 65.6 & $82.7^{+0.1}_{-0.2}$ & 0.000 & 0.000 & 0.000 \\
	Terzan 12 & BD & 73.4 & -60.3 & 1.7 & 89.0 & $-89.3^{+0.5}_{-0.8}$ & 0.000 & 0.000 & 0.000 \\
	NGC 6569 & OH & 187.1 & -64.6 & 3.6 & 59.4 & $63.1^{+2.7}_{-3.6}$ & 0.000 & 0.000 & 0.000 \\
	BH 261 & UN & 287.0 & -55.0 & 2.4 & 73.5 & $-77.0^{+1.6}_{-2.2}$ & 0.000 & 0.000 & 0.000 \\
	NGC 6584 & YH & 81.6 & -40.9 & 6.4 & 42.3 & $85.6^{+2.9}_{-2.8}$ & 0.000 & 0.004 & 0.000 \\
	NGC 6624 & BD & 256.4 & -15.3 & 17.2 & 21.0 & $82.9^{+4.9}_{-15.0}$ & 0.021 & 0.054 & 0.000 \\
	NGC 6626 & OH & 259.8 & -25.7 & 5.5 & 80.9 & $88.6^{+1.0}_{-1.8}$ & 0.000 & 0.000 & 0.000 \\
	NGC 6638 & OH & 309.6 & -13.3 & 8.3 & 22.7 & $-42.7^{+7.4}_{-6.0}$ & 0.169 & 0.000 & 0.000 \\
	NGC 6637 & BD & 246.3 & -16.0 & 5.7 & 13.6 & $76.8^{+2.4}_{-1.6}$ & 0.000 & 0.000 & 0.000 \\
	NGC 6642 & YH & 346.3 & -29.5 & 52.0 & 17.1 & $-49.3^{+17.5}_{-26.5}$ & 0.313 & 0.351 & 0.000 \\
	NGC 6652 & OH & 257.0 & -17.7 & 12.8 & 35.1 & $86.6^{+2.0}_{-2.0}$ & 0.005 & 0.009 & 0.001 \\
	NGC 6656 & OH & 270.6 & -55.8 & 1.7 & 85.2 & $-86.0^{+0.1}_{-0.1}$ & 0.000 & 0.000 & 0.000 \\
	Pal 8 & BD & 172.3 & -67.9 & 2.4 & 39.7 & $65.0^{+1.7}_{-1.9}$ & 0.000 & 0.000 & 0.000 \\
	NGC 6681 & OH & 292.4 & -4.6 & 6.3 & 12.6 & $-57.4^{+2.5}_{-3.4}$ & 0.001 & 0.001 & 0.000 \\
	NGC 6712 & OH & 213.9 & -6.9 & 8.7 & 43.5 & $44.4^{+6.7}_{-6.0}$ & 0.104 & 0.000 & 0.000 \\
	NGC 6715 & SG & 274.0 & -11.3 & 2.6 & 60.4 & $-76.2^{+1.2}_{-1.2}$ & 0.000 & 0.000 & 0.000 \\
	NGC 6717 & OH & 324.2 & -56.0 & 4.6 & 42.7 & $-62.4^{+1.8}_{-1.4}$ & 0.000 & 0.000 & 0.000 \\
	NGC 6723 & OH & 31.5 & 3.5 & 7.2 & 1.0 & $-42.1^{+2.6}_{-1.4}$ & 0.000 & 0.000 & 0.000 \\
	NGC 6749 & OH & 296.9 & -86.5 & 1.3 & 31.4 & $89.1^{+0.6}_{-0.9}$ & 0.000 & 0.000 & 0.000 \\
	NGC 6752 & OH & 58.7 & -65.3 & 1.3 & 58.8 & $-84.1^{+0.4}_{-0.3}$ & 0.000 & 0.000 & 0.000 \\
	NGC 6760 & BD & 275.8 & -83.7 & 1.1 & 37.2 & $89.0^{+0.7}_{-0.9}$ & 0.000 & 0.000 & 0.000 \\
	NGC 6779 & OH & 22.8 & 14.3 & 4.7 & 34.7 & $-35.1^{+1.8}_{-1.3}$ & 0.845 & 0.000 & 0.000 \\
	Terzan 7 & SG & 271.9 & -6.0 & 2.6 & 53.2 & $-77.7^{+1.5}_{-1.5}$ & 0.000 & 0.000 & 0.000 \\
	Pal 10 & BD & 63.1 & -82.1 & 0.6 & 54.4 & $89.4^{+0.4}_{-0.4}$ & 0.000 & 0.000 & 0.000 \\
	Arp 2 & SG & 274.5 & -13.8 & 2.9 & 51.7 & $-75.9^{+1.7}_{-1.8}$ & 0.000 & 0.000 & 0.000 \\
	NGC 6809 & OH & 270.2 & -19.9 & 3.6 & 60.0 & $-80.8^{+1.8}_{-1.5}$ & 0.000 & 0.000 & 0.000 \\
	Terzan 8 & SG & 273.3 & -7.8 & 2.6 & 48.3 & $-76.5^{+1.7}_{-1.8}$ & 0.000 & 0.000 & 0.000 \\
	Pal 11 & BD & 233.5 & -63.3 & 2.4 & 9.5 & $76.1^{+2.2}_{-2.0}$ & 0.000 & 0.000 & 0.000 \\
	NGC 6838 & BD & 48.8 & -77.8 & 0.7 & 71.4 & $-86.6^{+0.5}_{-0.6}$ & 0.000 & 0.000 & 0.000 \\
	NGC 6864 & OH & 159.8 & -36.2 & 18.2 & 31.5 & $35.4^{+6.1}_{-4.0}$ & 0.593 & 0.055 & 0.062 \\
	NGC 6934 & YH & 257.7 & -66.4 & 3.2 & 6.7 & $86.7^{+2.1}_{-2.5}$ & 0.000 & 0.000 & 0.000 \\
	NGC 6981 & YH & 285.2 & -37.4 & 82.9 & 6.7 & $-63.2^{+17.0}_{-31.2}$ & 0.187 & 0.569 & 0.000 \\
	NGC 7006 & YH & 4.9 & 41.1 & 23.7 & 2.6 & $-41.5^{+19.5}_{-19.8}$ & 0.405 & 0.095 & 0.000 \\
	NGC 7078 & YH & 272.2 & -60.8 & 3.1 & 34.4 & $-86.1^{+2.4}_{-2.6}$ & 0.000 & 0.000 & 0.000 \\
	NGC 7089 & OH & 170.2 & 29.5 & 18.1 & 24.1 & $32.9^{+4.0}_{-4.9}$ & 0.841 & 0.054 & 0.065 \\
	NGC 7099 & OH & 113.0 & 31.3 & 11.3 & 33.0 & $63.6^{+9.2}_{-9.9}$ & 0.005 & 0.001 & 0.000 \\
	Pal 12 & SG & 279.3 & -22.9 & 1.9 & 3.3 & $-72.8^{+1.8}_{-2.0}$ & 0.000 & 0.000 & 0.000 \\
	Pal 13 & YH & 46.5 & 26.8 & 6.6 & 25.2 & $-59.8^{+5.7}_{-5.6}$ & 0.000 & 0.000 & 0.000 \\
	NGC 7492 & OH & 178.4 & 3.5 & 7.9 & 9.2 & $11.8^{+4.4}_{-2.1}$ & 1.000 & 0.001 & 0.141 \\
	\bottomrule
\end{tabular}
\endgroup
\caption{Continuation of Table \ref{tab:GCprops-full}.}
\end{table*}

\begin{table*}
\scriptsize
\centering
\begin{tabular}{lcrrccccccccr}
	\toprule
	Name & Class & RA & Dec & Distance & $\Delta \theta$ & Length & Width & $l_\text{normal}$ & $b_\text{normal}$ & $\theta_\text{obs}$ & $p_\text{inVPOS}$ & Ref. \\
	 &  & [deg] & [deg] & [kpc] & [deg] & [deg] & [deg] & [deg] & [deg] & [deg] &  &  \\
	\midrule
	\multirow{2}{*}{20.0-1} & \multirow{2}{*}{3} & 112.92 & 61.56 & $12.6 \pm 1.3$ & $1.8$ & \multirow{2}{*}{158.6} & \multirow{2}{*}{1.8} & \multirow{2}{*}{270.2} & \multirow{2}{*}{38.0} & \multirow{2}{*}{$76.4^{+8.3}_{-6.0}$} & \multirow{2}{*}{0.000} & \multirow{2}{*}{\citet{Mateu:2018}} \\
	 &  & 273.92 & -43.35 & $28.1 \pm 2.8$ & $1.8$ &  &  &  &  &  &  &  \\
	\multirow{2}{*}{Acheron} & \multirow{2}{*}{1} & 230.00 & -2.00 & $3.8 \pm 0.8$ & $0.5$ & \multirow{2}{*}{36.5} & \multirow{2}{*}{0.4} & \multirow{2}{*}{228.2} & \multirow{2}{*}{-53.6} & \multirow{2}{*}{$70.4^{+6.9}_{-5.1}$} & \multirow{2}{*}{0.000} & \multirow{2}{*}{\citet{Grillmair:2009}} \\
	 &  & 259.00 & 21.00 & $3.5 \pm 0.7$ & $0.5$ &  &  &  &  &  &  &  \\
	\multirow{2}{*}{ACS} & \multirow{2}{*}{1} & 126.40 & -0.70 & $8.9 \pm 0.2$ & $2.1$ & \multirow{2}{*}{65.1} & \multirow{2}{*}{2.1} & \multirow{2}{*}{143.5} & \multirow{2}{*}{-68.0} & \multirow{2}{*}{$67.5^{+0.4}_{-0.4}$} & \multirow{2}{*}{0.000} & \multirow{2}{*}{\citet{Grillmair:2006a}} \\
	 &  & 133.90 & 64.20 & $8.9 \pm 0.2$ & $2.1$ &  &  &  &  &  &  &  \\
	\multirow{2}{*}{Aliqa Uma} & \multirow{2}{*}{1} & 31.70 & -31.50 & $28.8 \pm 5.8$ & $0.3$ & \multirow{2}{*}{10.0} & \multirow{2}{*}{0.3} & \multirow{2}{*}{171.3} & \multirow{2}{*}{23.8} & \multirow{2}{*}{$29.9^{+13.5}_{-5.3}$} & \multirow{2}{*}{0.717} & \multirow{2}{*}{\citet{Shipp:2018}} \\
	 &  & 40.60 & -38.30 & $28.8 \pm 5.8$ & $0.3$ &  &  &  &  &  &  &  \\
	\multirow{2}{*}{Alpheus} & \multirow{2}{*}{1} & 27.70 & -45.00 & $2.0 \pm 0.7$ & $1.4$ & \multirow{2}{*}{24.2} & \multirow{2}{*}{1.4} & \multirow{2}{*}{281.0} & \multirow{2}{*}{-24.1} & \multirow{2}{*}{$76.2^{+8.7}_{-4.6}$} & \multirow{2}{*}{0.000} & \multirow{2}{*}{\citet{Grillmair:2013}} \\
	 &  & 21.58 & -69.00 & $1.6 \pm 0.8$ & $1.4$ &  &  &  &  &  &  &  \\
	\multirow{2}{*}{ATLAS} & \multirow{2}{*}{1} & 9.30 & -20.90 & $22.9 \pm 4.6$ & $0.2$ & \multirow{2}{*}{22.6} & \multirow{2}{*}{0.2} & \multirow{2}{*}{157.4} & \multirow{2}{*}{21.3} & \multirow{2}{*}{$28.4^{+5.9}_{-3.5}$} & \multirow{2}{*}{0.909} & \multirow{2}{*}{\citet{Shipp:2018}} \\
	 &  & 30.70 & -33.20 & $22.9 \pm 4.6$ & $0.2$ &  &  &  &  &  &  &  \\
	\multirow{2}{*}{Cetus} & \multirow{2}{*}{3} & 20.01 & -4.12 & $32.5 \pm 1.3$ & $3.5$ & \multirow{2}{*}{30.1} & \multirow{2}{*}{9.4} & \multirow{2}{*}{243.1} & \multirow{2}{*}{1.7} & \multirow{2}{*}{$73.9^{+4.5}_{-4.5}$} & \multirow{2}{*}{0.000} & \multirow{2}{*}{\citet{Yam:2013}} \\
	 &  & 31.97 & 23.61 & $27.2 \pm 1.7$ & $4.7$ &  &  &  &  &  &  &  \\
	\multirow{2}{*}{Chenab} & \multirow{2}{*}{1} & 319.30 & -59.90 & $39.8 \pm 8.0$ & $0.7$ & \multirow{2}{*}{18.5} & \multirow{2}{*}{0.7} & \multirow{2}{*}{208.8} & \multirow{2}{*}{-22.1} & \multirow{2}{*}{$43.0^{+5.8}_{-6.5}$} & \multirow{2}{*}{0.170} & \multirow{2}{*}{\citet{Shipp:2018}} \\
	 &  & 331.70 & -43.00 & $39.8 \pm 8.0$ & $0.7$ &  &  &  &  &  &  &  \\
	\multirow{2}{*}{Cocytos} & \multirow{2}{*}{1} & 186.00 & -3.00 & $11.0 \pm 2.0$ & $0.5$ & \multirow{2}{*}{75.1} & \multirow{2}{*}{0.3} & \multirow{2}{*}{164.5} & \multirow{2}{*}{-25.8} & \multirow{2}{*}{$25.4^{+7.0}_{-6.3}$} & \multirow{2}{*}{0.931} & \multirow{2}{*}{\citet{Grillmair:2009}} \\
	 &  & 259.00 & 20.00 & $11.0 \pm 2.0$ & $0.5$ &  &  &  &  &  &  &  \\
	\multirow{2}{*}{Corvus} & \multirow{2}{*}{3} & 84.98 & -16.95 & $3.2 \pm 0.3$ & $1.6$ & \multirow{2}{*}{141.0} & \multirow{2}{*}{1.6} & \multirow{2}{*}{270.2} & \multirow{2}{*}{56.7} & \multirow{2}{*}{$77.5^{+8.0}_{-7.5}$} & \multirow{2}{*}{0.000} & \multirow{2}{*}{\citet{Mateu:2018}} \\
	 &  & 266.16 & -22.06 & $19.4 \pm 1.9$ & $1.6$ &  &  &  &  &  &  &  \\
	\multirow{2}{*}{EBS} & \multirow{2}{*}{3} & 132.25 & 13.60 & $9.4 \pm 1.4$ & $0.5$ & \multirow{2}{*}{13.8} & \multirow{2}{*}{0.2} & \multirow{2}{*}{181.0} & \multirow{2}{*}{-71.2} & \multirow{2}{*}{$68.7^{+5.8}_{-6.6}$} & \multirow{2}{*}{0.000} & \multirow{2}{*}{\citet{Grillmair:2011}} \\
	 &  & 135.40 & 0.12 & $9.7 \pm 0.9$ & $0.5$ &  &  &  &  &  &  &  \\
	\multirow{2}{*}{Elqui} & \multirow{2}{*}{1} & 10.70 & -36.90 & $50.1 \pm 10.0$ & $0.5$ & \multirow{2}{*}{9.4} & \multirow{2}{*}{0.5} & \multirow{2}{*}{159.6} & \multirow{2}{*}{0.0} & \multirow{2}{*}{$14.1^{+12.7}_{-7.5}$} & \multirow{2}{*}{0.959} & \multirow{2}{*}{\citet{Shipp:2018}} \\
	 &  & 20.60 & -42.40 & $50.1 \pm 10.0$ & $0.5$ &  &  &  &  &  &  &  \\
	\multirow{2}{*}{Fimbulthul} & \multirow{2}{*}{2} & 198.74 & -29.56 & $4.2 \pm 0.0$ & $0.5$ & \multirow{2}{*}{15.5} & \multirow{2}{*}{--} & \multirow{2}{*}{135.5} & \multirow{2}{*}{-44.5} & \multirow{2}{*}{$51.2^{+0.3}_{-0.3}$} & \multirow{2}{*}{0.000} & \multirow{2}{*}{\citet{Ibata:2019}} \\
	 &  & 214.23 & -22.76 & $4.2 \pm 0.0$ & $0.5$ &  &  &  &  &  &  &  \\
	\multirow{2}{*}{Fj\"{o}rm} & \multirow{2}{*}{2} & 197.37 & 5.55 & $4.9 \pm 0.1$ & $0.5$ & \multirow{2}{*}{69.8} & \multirow{2}{*}{--} & \multirow{2}{*}{219.7} & \multirow{2}{*}{-53.1} & \multirow{2}{*}{$65.1^{+0.3}_{-0.3}$} & \multirow{2}{*}{0.000} & \multirow{2}{*}{\citet{Ibata:2019}} \\
	 &  & 250.88 & 64.20 & $4.9 \pm 0.1$ & $0.5$ &  &  &  &  &  &  &  \\
	\multirow{2}{*}{Gaia-1} & \multirow{2}{*}{1} & 184.00 & -18.00 & $6.0 \pm 0.6$ & $0.5$ & \multirow{2}{*}{20.5} & \multirow{2}{*}{0.5} & \multirow{2}{*}{171.5} & \multirow{2}{*}{-54.6} & \multirow{2}{*}{$51.9^{+9.3}_{-7.1}$} & \multirow{2}{*}{0.001} & \multirow{2}{*}{\citet{Malhan:2018}} \\
	 &  & 197.00 & -2.00 & $5.0 \pm 0.5$ & $0.5$ &  &  &  &  &  &  &  \\
	\multirow{2}{*}{Gaia-2} & \multirow{2}{*}{1} & 6.00 & -22.00 & $13.0 \pm 1.3$ & $0.5$ & \multirow{2}{*}{9.6} & \multirow{2}{*}{--} & \multirow{2}{*}{120.3} & \multirow{2}{*}{22.4} & \multirow{2}{*}{$54.0^{+6.1}_{-8.6}$} & \multirow{2}{*}{0.004} & \multirow{2}{*}{\citet{Malhan:2018}} \\
	 &  & 15.00 & -27.00 & $10.0 \pm 1.0$ & $0.5$ &  &  &  &  &  &  &  \\
	\multirow{2}{*}{Gaia-3} & \multirow{2}{*}{1} & 171.00 & -15.00 & $9.0 \pm 0.9$ & $0.5$ & \multirow{2}{*}{18.5} & \multirow{2}{*}{--} & \multirow{2}{*}{228.2} & \multirow{2}{*}{-58.6} & \multirow{2}{*}{$71.7^{+3.6}_{-4.4}$} & \multirow{2}{*}{0.000} & \multirow{2}{*}{\citet{Malhan:2018}} \\
	 &  & 179.00 & -32.00 & $14.0 \pm 1.4$ & $0.5$ &  &  &  &  &  &  &  \\
	\multirow{2}{*}{Gaia-4} & \multirow{2}{*}{1} & 163.00 & -11.00 & $11.5 \pm 1.1$ & $0.5$ & \multirow{2}{*}{8.9} & \multirow{2}{*}{--} & \multirow{2}{*}{164.4} & \multirow{2}{*}{-41.0} & \multirow{2}{*}{$39.7^{+14.5}_{-7.6}$} & \multirow{2}{*}{0.398} & \multirow{2}{*}{\citet{Malhan:2018}} \\
	 &  & 167.00 & -3.00 & $10.7 \pm 1.1$ & $0.5$ &  &  &  &  &  &  &  \\
	\multirow{2}{*}{Gaia-5} & \multirow{2}{*}{1} & 137.00 & 23.00 & $20.5 \pm 2.0$ & $0.5$ & \multirow{2}{*}{23.7} & \multirow{2}{*}{--} & \multirow{2}{*}{137.6} & \multirow{2}{*}{-42.8} & \multirow{2}{*}{$49.1^{+0.9}_{-0.8}$} & \multirow{2}{*}{0.000} & \multirow{2}{*}{\citet{Malhan:2018}} \\
	 &  & 154.00 & 42.00 & $18.5 \pm 1.9$ & $0.5$ &  &  &  &  &  &  &  \\
	\multirow{2}{*}{GD-1} & \multirow{2}{*}{1} & 135.00 & 17.00 & $6.5 \pm 0.7$ & $0.5$ & \multirow{2}{*}{57.4} & \multirow{2}{*}{0.2} & \multirow{2}{*}{120.8} & \multirow{2}{*}{-47.1} & \multirow{2}{*}{$61.0^{+1.0}_{-1.0}$} & \multirow{2}{*}{0.000} & \multirow{2}{*}{\citet{Malhan:2018}} \\
	 &  & 190.00 & 58.00 & $10.0 \pm 1.0$ & $0.5$ &  &  &  &  &  &  &  \\
	\multirow{2}{*}{Gj\"{o}ll} & \multirow{2}{*}{2} & 70.16 & -2.46 & $3.4 \pm 0.1$ & $0.5$ & \multirow{2}{*}{26.3} & \multirow{2}{*}{--} & \multirow{2}{*}{299.2} & \multirow{2}{*}{-69.6} & \multirow{2}{*}{$79.8^{+0.7}_{-0.8}$} & \multirow{2}{*}{0.000} & \multirow{2}{*}{\citet{Ibata:2019}} \\
	 &  & 90.05 & -20.15 & $3.4 \pm 0.1$ & $0.5$ &  &  &  &  &  &  &  \\
	\multirow{2}{*}{Hermus} & \multirow{2}{*}{3} & 245.40 & 5.00 & $19.0 \pm 3.0$ & $0.5$ & \multirow{2}{*}{45.4} & \multirow{2}{*}{0.3} & \multirow{2}{*}{238.0} & \multirow{2}{*}{35.1} & \multirow{2}{*}{$74.4^{+3.5}_{-4.6}$} & \multirow{2}{*}{0.001} & \multirow{2}{*}{\citet{Grillmair:2014}} \\
	 &  & 253.05 & 49.90 & $15.0 \pm 3.0$ & $0.5$ &  &  &  &  &  &  &  \\
	\multirow{2}{*}{Hyllus} & \multirow{2}{*}{3} & 249.00 & 11.00 & $23.0 \pm 3.0$ & $0.5$ & \multirow{2}{*}{23.1} & \multirow{2}{*}{0.2} & \multirow{2}{*}{253.2} & \multirow{2}{*}{39.3} & \multirow{2}{*}{$87.0^{+2.0}_{-2.7}$} & \multirow{2}{*}{0.000} & \multirow{2}{*}{\citet{Grillmair:2014}} \\
	 &  & 246.90 & 34.00 & $18.5 \pm 3.0$ & $0.5$ &  &  &  &  &  &  &  \\
	\multirow{2}{*}{Indus} & \multirow{2}{*}{1} & 323.70 & -50.70 & $16.6 \pm 3.3$ & $0.8$ & \multirow{2}{*}{20.3} & \multirow{2}{*}{0.8} & \multirow{2}{*}{141.2} & \multirow{2}{*}{-18.5} & \multirow{2}{*}{$33.3^{+12.1}_{-18.0}$} & \multirow{2}{*}{0.598} & \multirow{2}{*}{\citet{Shipp:2018}} \\
	 &  & 352.00 & -64.80 & $16.6 \pm 3.3$ & $0.8$ &  &  &  &  &  &  &  \\
	\multirow{2}{*}{Jet} & \multirow{2}{*}{1} & 134.67 & -26.58 & $28.6 \pm 0.8$ & $0.2$ & \multirow{2}{*}{11.5} & \multirow{2}{*}{0.2} & \multirow{2}{*}{150.6} & \multirow{2}{*}{-14.8} & \multirow{2}{*}{$22.0^{+0.9}_{-0.6}$} & \multirow{2}{*}{1.000} & \multirow{2}{*}{\citet{Jethwa:2018}} \\
	 &  & 142.33 & -17.53 & $28.6 \pm 0.8$ & $0.2$ &  &  &  &  &  &  &  \\
	\multirow{2}{*}{Jhelum} & \multirow{2}{*}{1} & 321.20 & -45.10 & $13.2 \pm 2.6$ & $1.2$ & \multirow{2}{*}{29.2} & \multirow{2}{*}{1.2} & \multirow{2}{*}{298.4} & \multirow{2}{*}{7.4} & \multirow{2}{*}{$51.2^{+8.5}_{-17.8}$} & \multirow{2}{*}{0.186} & \multirow{2}{*}{\citet{Shipp:2018}} \\
	 &  & 4.70 & -51.70 & $13.2 \pm 2.6$ & $1.2$ &  &  &  &  &  &  &  \\
	\multirow{2}{*}{Kshir} & \multirow{2}{*}{3} & 230.38 & 68.17 & $10.0 \pm 1.0$ & $0.5$ & \multirow{2}{*}{62.1} & \multirow{2}{*}{--} & \multirow{2}{*}{532.3} & \multirow{2}{*}{-57.5} & \multirow{2}{*}{$54.9^{+1.4}_{-1.2}$} & \multirow{2}{*}{0.000} & \multirow{2}{*}{\citet{Malhan:2019}} \\
	 &  & 153.05 & 25.11 & $10.0 \pm 1.0$ & $0.5$ &  &  &  &  &  &  &  \\
	\multirow{2}{*}{Kwando} & \multirow{2}{*}{3} & 19.00 & -23.90 & $20.3 \pm 4.1$ & $0.5$ & \multirow{2}{*}{12.0} & \multirow{2}{*}{0.2} & \multirow{2}{*}{151.6} & \multirow{2}{*}{24.3} & \multirow{2}{*}{$36.2^{+11.4}_{-5.4}$} & \multirow{2}{*}{0.535} & \multirow{2}{*}{\citet{Grillmair:2017b}} \\
	 &  & 31.00 & -29.33 & $20.3 \pm 4.1$ & $0.5$ &  &  &  &  &  &  &  \\
	\multirow{2}{*}{Leiptr} & \multirow{2}{*}{2} & 61.03 & 0.72 & $7.9 \pm 0.4$ & $0.5$ & \multirow{2}{*}{48.0} & \multirow{2}{*}{--} & \multirow{2}{*}{151.3} & \multirow{2}{*}{69.2} & \multirow{2}{*}{$73.0^{+0.9}_{-0.8}$} & \multirow{2}{*}{0.000} & \multirow{2}{*}{\citet{Ibata:2019}} \\
	 &  & 96.18 & -34.21 & $7.9 \pm 0.4$ & $0.5$ &  &  &  &  &  &  &  \\
	\multirow{2}{*}{Lethe} & \multirow{2}{*}{1} & 258.00 & 20.00 & $13.4 \pm 2.7$ & $0.5$ & \multirow{2}{*}{81.2} & \multirow{2}{*}{0.2} & \multirow{2}{*}{504.1} & \multirow{2}{*}{-30.4} & \multirow{2}{*}{$37.1^{+4.8}_{-4.0}$} & \multirow{2}{*}{0.477} & \multirow{2}{*}{\citet{Grillmair:2009}} \\
	 &  & 171.00 & 18.00 & $12.2 \pm 2.4$ & $0.5$ &  &  &  &  &  &  &  \\
	\multirow{2}{*}{Magellanic} & \multirow{2}{*}{3} & 355.93 & -11.91 & $55.0 \pm 11.0$ & $2.0$ & \multirow{2}{*}{17.8} & \multirow{2}{*}{--} & \multirow{2}{*}{179.3} & \multirow{2}{*}{2.9} & \multirow{2}{*}{$12.4^{+8.2}_{-4.1}$} & \multirow{2}{*}{0.988} & \multirow{2}{*}{\citet{Bruns:2005}} \\
	 &  & 5.44 & -27.34 & $55.0 \pm 11.0$ & $2.0$ &  &  &  &  &  &  &  \\
	\multirow{2}{*}{Molonglo} & \multirow{2}{*}{1} & 6.40 & -24.40 & $22.9 \pm 4.6$ & $0.3$ & \multirow{2}{*}{7.4} & \multirow{2}{*}{0.3} & \multirow{2}{*}{152.3} & \multirow{2}{*}{17.1} & \multirow{2}{*}{$35.5^{+17.3}_{-12.2}$} & \multirow{2}{*}{0.528} & \multirow{2}{*}{\citet{Shipp:2018}} \\
	 &  & 13.60 & -28.10 & $22.9 \pm 4.6$ & $0.3$ &  &  &  &  &  &  &  \\
	\multirow{2}{*}{Murrumbidgee} & \multirow{2}{*}{3} & 358.71 & 15.97 & $20.0 \pm 4.0$ & $0.5$ & \multirow{2}{*}{43.8} & \multirow{2}{*}{0.2} & \multirow{2}{*}{201.4} & \multirow{2}{*}{22.0} & \multirow{2}{*}{$40.1^{+5.2}_{-3.3}$} & \multirow{2}{*}{0.162} & \multirow{2}{*}{\citet{Grillmair:2017b}} \\
	 &  & 13.90 & -25.27 & $20.0 \pm 4.0$ & $0.5$ &  &  &  &  &  &  &  \\
	\multirow{2}{*}{NGC 5053} & \multirow{2}{*}{3} & 199.11 & 17.70 & $17.4 \pm 0.8$ & $0.0$ & \multirow{2}{*}{4.8} & \multirow{2}{*}{--} & \multirow{2}{*}{517.9} & \multirow{2}{*}{-13.7} & \multirow{2}{*}{$39.2^{+20.2}_{-21.3}$} & \multirow{2}{*}{0.467} & \multirow{2}{*}{\citet{Lauchner:2006}} \\
	 &  & 195.00 & 15.00 & $17.4 \pm 3.5$ & $1.0$ &  &  &  &  &  &  &  \\
	\multirow{2}{*}{NGC 5466} & \multirow{2}{*}{3} & 211.37 & 28.53 & $16.0 \pm 0.7$ & $0.0$ & \multirow{2}{*}{28.7} & \multirow{2}{*}{0.6} & \multirow{2}{*}{276.4} & \multirow{2}{*}{13.0} & \multirow{2}{*}{$72.9^{+2.8}_{-3.0}$} & \multirow{2}{*}{0.000} & \multirow{2}{*}{\citet{Grillmair:2006b}} \\
	 &  & 180.00 & 42.00 & $16.0 \pm 3.2$ & $2.0$ &  &  &  &  &  &  &  \\
	\bottomrule
\end{tabular}

\caption{The full list of streams analyzed in this work.
For column descriptions, see Table \ref{tab:streamprops}.}
\label{tab:streamprops-full}
\end{table*}

\begin{table*}
\scriptsize
\centering
\begin{tabular}{lcrrccccccccr}
	\toprule
	Name & Class & RA & Dec & Distance & $\Delta \theta$ & Length & Width & $l_\text{normal}$ & $b_\text{normal}$ & $\theta_\text{obs}$ & $p_\text{inVPOS}$ & Ref. \\
	 &  & [deg] & [deg] & [kpc] & [deg] & [deg] & [deg] & [deg] & [deg] & [deg] &  &  \\
	\midrule
	\multirow{2}{*}{NGC 5904} & \multirow{2}{*}{3} & 229.64 & 2.08 & $7.5 \pm 0.3$ & $0.0$ & \multirow{2}{*}{42.7} & \multirow{2}{*}{0.7} & \multirow{2}{*}{290.1} & \multirow{2}{*}{13.9} & \multirow{2}{*}{$59.4^{+6.4}_{-9.2}$} & \multirow{2}{*}{0.018} & \multirow{2}{*}{\citet{Grillmair:2019}} \\
	 &  & 190.00 & 20.26 & $7.5 \pm 1.5$ & $1.0$ &  &  &  &  &  &  &  \\
	\multirow{2}{*}{Orinoco} & \multirow{2}{*}{3} & 0.00 & -25.51 & $20.0 \pm 3.0$ & $0.5$ & \multirow{2}{*}{20.7} & \multirow{2}{*}{0.3} & \multirow{2}{*}{131.3} & \multirow{2}{*}{14.8} & \multirow{2}{*}{$41.8^{+6.6}_{-8.4}$} & \multirow{2}{*}{0.267} & \multirow{2}{*}{\citet{Grillmair:2017b}} \\
	 &  & 22.97 & -28.61 & $20.0 \pm 3.0$ & $0.5$ &  &  &  &  &  &  &  \\
	\multirow{2}{*}{Orphan} & \multirow{2}{*}{1} & 145.00 & 40.00 & $38.0 \pm 3.8$ & $0.5$ & \multirow{2}{*}{21.1} & \multirow{2}{*}{2.0} & \multirow{2}{*}{210.7} & \multirow{2}{*}{-45.4} & \multirow{2}{*}{$55.9^{+2.2}_{-2.3}$} & \multirow{2}{*}{0.000} & \multirow{2}{*}{\citet{Malhan:2018}} \\
	 &  & 153.00 & 20.00 & $33.0 \pm 3.3$ & $0.5$ &  &  &  &  &  &  &  \\
	\multirow{2}{*}{Pal 5} & \multirow{2}{*}{3} & 219.36 & -12.02 & $23.2 \pm 2.3$ & $0.5$ & \multirow{2}{*}{27.0} & \multirow{2}{*}{0.3} & \multirow{2}{*}{172.4} & \multirow{2}{*}{28.2} & \multirow{2}{*}{$31.4^{+1.7}_{-1.8}$} & \multirow{2}{*}{1.000} & \multirow{2}{*}{\citet{Starkman:2019}} \\
	 &  & 239.69 & 5.84 & $23.2 \pm 2.3$ & $0.5$ &  &  &  &  &  &  &  \\
	\multirow{2}{*}{Palca} & \multirow{2}{*}{1} & 30.30 & -53.70 & $36.3 \pm 7.3$ & $0.8$ & \multirow{2}{*}{57.3} & \multirow{2}{*}{0.8} & \multirow{2}{*}{205.9} & \multirow{2}{*}{20.4} & \multirow{2}{*}{$42.7^{+3.0}_{-2.7}$} & \multirow{2}{*}{0.015} & \multirow{2}{*}{\citet{Shipp:2018}} \\
	 &  & 16.20 & 2.40 & $36.3 \pm 7.3$ & $0.8$ &  &  &  &  &  &  &  \\
	\multirow{2}{*}{PAndAS} & \multirow{2}{*}{3} & 4.95 & 45.92 & $17.0 \pm 3.0$ & $0.5$ & \multirow{2}{*}{12.3} & \multirow{2}{*}{1.0} & \multirow{2}{*}{196.8} & \multirow{2}{*}{65.8} & \multirow{2}{*}{$71.5^{+4.7}_{-6.0}$} & \multirow{2}{*}{0.000} & \multirow{2}{*}{\citet{Martin:2014}} \\
	 &  & 20.47 & 40.97 & $17.0 \pm 3.0$ & $0.5$ &  &  &  &  &  &  &  \\
	\multirow{2}{*}{Pegasus} & \multirow{2}{*}{1} & 328.30 & 20.80 & $18.0 \pm 2.0$ & $0.2$ & \multirow{2}{*}{8.6} & \multirow{2}{*}{0.2} & \multirow{2}{*}{233.0} & \multirow{2}{*}{-52.3} & \multirow{2}{*}{$72.4^{+6.4}_{-15.3}$} & \multirow{2}{*}{0.041} & \multirow{2}{*}{\citet{Perottoni:2019}} \\
	 &  & 333.40 & 28.10 & $18.0 \pm 2.0$ & $0.2$ &  &  &  &  &  &  &  \\
	\multirow{2}{*}{Phlegethon} & \multirow{2}{*}{2} & 284.75 & -59.20 & $3.6 \pm 0.3$ & $1.4$ & \multirow{2}{*}{66.6} & \multirow{2}{*}{1.4} & \multirow{2}{*}{526.4} & \multirow{2}{*}{72.9} & \multirow{2}{*}{$75.8^{+1.2}_{-1.3}$} & \multirow{2}{*}{0.000} & \multirow{2}{*}{\citet{Ibata:2018}} \\
	 &  & 324.50 & -0.26 & $3.6 \pm 0.3$ & $1.4$ &  &  &  &  &  &  &  \\
	\multirow{2}{*}{Phoenix} & \multirow{2}{*}{1} & 20.10 & -55.30 & $19.1 \pm 3.8$ & $0.2$ & \multirow{2}{*}{13.6} & \multirow{2}{*}{0.2} & \multirow{2}{*}{235.6} & \multirow{2}{*}{29.2} & \multirow{2}{*}{$71.4^{+9.9}_{-23.3}$} & \multirow{2}{*}{0.098} & \multirow{2}{*}{\citet{Shipp:2018}} \\
	 &  & 27.90 & -42.70 & $19.1 \pm 3.8$ & $0.2$ &  &  &  &  &  &  &  \\
	\multirow{2}{*}{PS1-A} & \multirow{2}{*}{2} & 28.39 & -6.55 & $7.9 \pm 1.9$ & $0.5$ & \multirow{2}{*}{4.9} & \multirow{2}{*}{0.2} & \multirow{2}{*}{249.3} & \multirow{2}{*}{22.9} & \multirow{2}{*}{$81.4^{+5.8}_{-14.7}$} & \multirow{2}{*}{0.000} & \multirow{2}{*}{\citet{Bernard:2016}} \\
	 &  & 30.14 & -1.97 & $7.9 \pm 1.9$ & $0.5$ &  &  &  &  &  &  &  \\
	\multirow{2}{*}{PS1-B} & \multirow{2}{*}{2} & 145.59 & -15.28 & $14.5 \pm 3.4$ & $0.5$ & \multirow{2}{*}{9.9} & \multirow{2}{*}{0.2} & \multirow{2}{*}{144.0} & \multirow{2}{*}{-24.2} & \multirow{2}{*}{$42.1^{+16.9}_{-9.6}$} & \multirow{2}{*}{0.348} & \multirow{2}{*}{\citet{Bernard:2016}} \\
	 &  & 151.13 & -7.00 & $14.5 \pm 3.4$ & $0.5$ &  &  &  &  &  &  &  \\
	\multirow{2}{*}{PS1-C} & \multirow{2}{*}{2} & 330.26 & 11.75 & $17.4 \pm 4.0$ & $0.5$ & \multirow{2}{*}{7.9} & \multirow{2}{*}{0.1} & \multirow{2}{*}{238.2} & \multirow{2}{*}{-46.1} & \multirow{2}{*}{$75.6^{+9.3}_{-24.8}$} & \multirow{2}{*}{0.072} & \multirow{2}{*}{\citet{Bernard:2016}} \\
	 &  & 334.98 & 18.20 & $17.4 \pm 4.0$ & $0.5$ &  &  &  &  &  &  &  \\
	\multirow{2}{*}{PS1-D} & \multirow{2}{*}{2} & 138.70 & -21.58 & $22.9 \pm 5.3$ & $0.5$ & \multirow{2}{*}{44.9} & \multirow{2}{*}{0.4} & \multirow{2}{*}{164.8} & \multirow{2}{*}{-52.6} & \multirow{2}{*}{$49.9^{+3.8}_{-3.1}$} & \multirow{2}{*}{0.000} & \multirow{2}{*}{\citet{Bernard:2016}} \\
	 &  & 140.74 & 23.28 & $22.9 \pm 5.3$ & $0.5$ &  &  &  &  &  &  &  \\
	\multirow{2}{*}{PS1-E} & \multirow{2}{*}{2} & 160.11 & 46.13 & $12.6 \pm 3.0$ & $0.5$ & \multirow{2}{*}{24.9} & \multirow{2}{*}{0.3} & \multirow{2}{*}{158.4} & \multirow{2}{*}{-52.8} & \multirow{2}{*}{$52.5^{+5.1}_{-3.4}$} & \multirow{2}{*}{0.000} & \multirow{2}{*}{\citet{Bernard:2016}} \\
	 &  & 192.94 & 62.89 & $12.6 \pm 3.0$ & $0.5$ &  &  &  &  &  &  &  \\
	\multirow{2}{*}{Ravi} & \multirow{2}{*}{1} & 334.80 & -44.10 & $22.9 \pm 4.6$ & $0.7$ & \multirow{2}{*}{16.6} & \multirow{2}{*}{0.7} & \multirow{2}{*}{170.6} & \multirow{2}{*}{-14.9} & \multirow{2}{*}{$18.1^{+11.3}_{-5.1}$} & \multirow{2}{*}{0.930} & \multirow{2}{*}{\citet{Shipp:2018}} \\
	 &  & 344.00 & -59.70 & $22.9 \pm 4.6$ & $0.7$ &  &  &  &  &  &  &  \\
	\multirow{2}{*}{Sangarius} & \multirow{2}{*}{3} & 134.06 & -17.79 & $21.0 \pm 5.0$ & $0.5$ & \multirow{2}{*}{50.5} & \multirow{2}{*}{0.4} & \multirow{2}{*}{164.2} & \multirow{2}{*}{-60.6} & \multirow{2}{*}{$57.8^{+3.0}_{-2.1}$} & \multirow{2}{*}{0.000} & \multirow{2}{*}{\citet{Grillmair:2017a}} \\
	 &  & 131.57 & 32.65 & $21.0 \pm 5.0$ & $0.5$ &  &  &  &  &  &  &  \\
	\multirow{2}{*}{Scamander} & \multirow{2}{*}{3} & 151.64 & -20.47 & $21.0 \pm 5.0$ & $0.5$ & \multirow{2}{*}{65.2} & \multirow{2}{*}{0.4} & \multirow{2}{*}{187.4} & \multirow{2}{*}{-54.9} & \multirow{2}{*}{$54.1^{+3.4}_{-2.5}$} & \multirow{2}{*}{0.000} & \multirow{2}{*}{\citet{Grillmair:2017a}} \\
	 &  & 143.61 & 44.34 & $21.0 \pm 5.0$ & $0.5$ &  &  &  &  &  &  &  \\
	\multirow{2}{*}{Slidr} & \multirow{2}{*}{2} & 148.38 & 14.99 & $3.6 \pm 0.1$ & $0.5$ & \multirow{2}{*}{31.6} & \multirow{2}{*}{--} & \multirow{2}{*}{273.1} & \multirow{2}{*}{-21.5} & \multirow{2}{*}{$78.3^{+0.7}_{-0.8}$} & \multirow{2}{*}{0.000} & \multirow{2}{*}{\citet{Ibata:2019}} \\
	 &  & 178.01 & 2.98 & $3.6 \pm 0.1$ & $0.5$ &  &  &  &  &  &  &  \\
	\multirow{2}{*}{Styx} & \multirow{2}{*}{1} & 259.00 & 21.00 & $50.0 \pm 10.0$ & $1.4$ & \multirow{2}{*}{60.4} & \multirow{2}{*}{1.4} & \multirow{2}{*}{499.6} & \multirow{2}{*}{-2.1} & \multirow{2}{*}{$29.7^{+2.3}_{-2.7}$} & \multirow{2}{*}{0.996} & \multirow{2}{*}{\citet{Grillmair:2009}} \\
	 &  & 194.00 & 20.00 & $38.0 \pm 7.6$ & $1.4$ &  &  &  &  &  &  &  \\
	\multirow{2}{*}{Sv\"{o}l} & \multirow{2}{*}{2} & 244.44 & 23.37 & $7.8 \pm 0.2$ & $0.5$ & \multirow{2}{*}{17.9} & \multirow{2}{*}{--} & \multirow{2}{*}{230.3} & \multirow{2}{*}{3.9} & \multirow{2}{*}{$61.2^{+4.7}_{-5.5}$} & \multirow{2}{*}{0.000} & \multirow{2}{*}{\citet{Ibata:2019}} \\
	 &  & 248.73 & 40.94 & $7.8 \pm 0.2$ & $0.5$ &  &  &  &  &  &  &  \\
	\multirow{2}{*}{Sylgr} & \multirow{2}{*}{2} & 167.79 & -8.35 & $4.1 \pm 0.0$ & $0.5$ & \multirow{2}{*}{20.0} & \multirow{2}{*}{--} & \multirow{2}{*}{299.8} & \multirow{2}{*}{29.9} & \multirow{2}{*}{$54.1^{+0.2}_{-0.2}$} & \multirow{2}{*}{0.000} & \multirow{2}{*}{\citet{Ibata:2019}} \\
	 &  & 186.58 & -1.26 & $4.1 \pm 0.0$ & $0.5$ &  &  &  &  &  &  &  \\
	\multirow{2}{*}{Triangulum/Pisces} & \multirow{2}{*}{1} & 21.35 & 34.98 & $26.0 \pm 4.0$ & $0.5$ & \multirow{2}{*}{12.0} & \multirow{2}{*}{0.2} & \multirow{2}{*}{222.2} & \multirow{2}{*}{24.6} & \multirow{2}{*}{$58.5^{+2.8}_{-1.6}$} & \multirow{2}{*}{0.000} & \multirow{2}{*}{\citet{Bonaca:2012}} \\
	 &  & 23.98 & 23.20 & $26.0 \pm 4.0$ & $0.5$ &  &  &  &  &  &  &  \\
	\multirow{2}{*}{Tucana III} & \multirow{2}{*}{1} & 353.70 & -59.70 & $25.1 \pm 5.0$ & $0.2$ & \multirow{2}{*}{4.8} & \multirow{2}{*}{0.2} & \multirow{2}{*}{286.0} & \multirow{2}{*}{25.4} & \multirow{2}{*}{$69.0^{+5.9}_{-18.4}$} & \multirow{2}{*}{0.095} & \multirow{2}{*}{\citet{Shipp:2018}} \\
	 &  & 3.20 & -59.40 & $25.1 \pm 5.0$ & $0.2$ &  &  &  &  &  &  &  \\
	\multirow{2}{*}{Turbio} & \multirow{2}{*}{1} & 28.00 & -61.00 & $16.6 \pm 3.3$ & $0.2$ & \multirow{2}{*}{15.0} & \multirow{2}{*}{0.2} & \multirow{2}{*}{209.0} & \multirow{2}{*}{32.4} & \multirow{2}{*}{$52.5^{+16.2}_{-22.7}$} & \multirow{2}{*}{0.248} & \multirow{2}{*}{\citet{Shipp:2018}} \\
	 &  & 27.90 & -46.00 & $16.6 \pm 3.3$ & $0.2$ &  &  &  &  &  &  &  \\
	\multirow{2}{*}{Turranburra} & \multirow{2}{*}{1} & 59.30 & -18.00 & $27.5 \pm 5.5$ & $0.6$ & \multirow{2}{*}{16.9} & \multirow{2}{*}{0.6} & \multirow{2}{*}{155.4} & \multirow{2}{*}{42.5} & \multirow{2}{*}{$47.0^{+5.7}_{-3.3}$} & \multirow{2}{*}{0.000} & \multirow{2}{*}{\citet{Shipp:2018}} \\
	 &  & 75.20 & -26.40 & $27.5 \pm 5.5$ & $0.6$ &  &  &  &  &  &  &  \\
	\multirow{2}{*}{Wambelong} & \multirow{2}{*}{1} & 90.50 & -45.60 & $15.1 \pm 3.0$ & $0.4$ & \multirow{2}{*}{14.2} & \multirow{2}{*}{0.4} & \multirow{2}{*}{183.5} & \multirow{2}{*}{61.9} & \multirow{2}{*}{$65.1^{+10.3}_{-17.3}$} & \multirow{2}{*}{0.007} & \multirow{2}{*}{\citet{Shipp:2018}} \\
	 &  & 79.30 & -34.30 & $15.1 \pm 3.0$ & $0.4$ &  &  &  &  &  &  &  \\
	\multirow{2}{*}{Willka Yaku} & \multirow{2}{*}{1} & 36.10 & -64.60 & $34.7 \pm 6.9$ & $0.2$ & \multirow{2}{*}{6.4} & \multirow{2}{*}{0.2} & \multirow{2}{*}{229.9} & \multirow{2}{*}{33.0} & \multirow{2}{*}{$67.8^{+12.1}_{-29.9}$} & \multirow{2}{*}{0.152} & \multirow{2}{*}{\citet{Shipp:2018}} \\
	 &  & 38.40 & -58.30 & $34.7 \pm 6.9$ & $0.2$ &  &  &  &  &  &  &  \\
	\multirow{2}{*}{Ylgr} & \multirow{2}{*}{2} & 169.05 & -10.36 & $9.5 \pm 0.3$ & $0.5$ & \multirow{2}{*}{29.4} & \multirow{2}{*}{--} & \multirow{2}{*}{184.2} & \multirow{2}{*}{-52.0} & \multirow{2}{*}{$50.8^{+2.3}_{-2.5}$} & \multirow{2}{*}{0.000} & \multirow{2}{*}{\citet{Ibata:2019}} \\
	 &  & 182.65 & -37.12 & $9.5 \pm 0.3$ & $0.5$ &  &  &  &  &  &  &  \\
	\bottomrule
\end{tabular}

\caption{Continuation of Table \ref{tab:streamprops-full}.}
\end{table*}

\bsp	
\label{lastpage}
\end{document}